\documentclass[a4paper]{article}
\pdfoutput=1

\usepackage{graphicx}

\usepackage{epsfig}
\usepackage{hyperref}
\usepackage{a4}
\usepackage{latexsym, amssymb} 
\usepackage{amsmath}
\usepackage{comment}
\usepackage{cite}

\usepackage{datetime}

\usepackage{booktabs}
\usepackage{multirow}

\newcommand{\lapprox}{%
\mathrel{%
\setbox0=\hbox{$<$}
\raise0.6ex\copy0\kern-\wd0
\lower0.65ex\hbox{$\sim$}
}}
\newcommand{\gapprox}{%
\mathrel{%
\setbox0=\hbox{$>$}
\raise0.6ex\copy0\kern-\wd0
\lower0.65ex\hbox{$\sim$}
}}

\catcode`@=11 
\def \gsim{\mathrel{\mathpalette\@versim>}}
\def \lsim{\mathrel{\mathpalette\@versim<}}
\def \@versim#1#2{\lower0.4ex\vbox{\baselineskip\z@skip\lineskip\z@skip
     \lineskiplimit\z@\ialign{$\m@th#1\hfil##\hfil$%
     \crcr#2\crcr\sim\crcr}}}
\catcode`@=12 

\makeatletter
 
 \@addtoreset{equation}{section}
\makeatother

\begin{document}

\hspace*{10cm}

\begin{center}

{\Large\bf Polarisations of the $Z$ and $W$ bosons in the processes $pp \to ZH$ and $pp \to W^{\pm}_{}H$}\\[15mm]

{\bf Junya Nakamura}
\\                    
\vskip .1cm
{Institut f\"ur theoretische Physik, Universit\"at T\"ubingen, Germany} \\
\vskip .2cm
{\small junya.nakamura@itp.uni-tuebingen.de} \\

\end{center}

\vskip 1.5cm

\begin{abstract}

The $Z$ boson in the process $pp \to ZH$ and the $W^+_{}$ and $W^-_{}$ in the process $pp \to W^{\pm}_{}H$ can be in polarised states. The polarisation density matrix of the $Z$ ($W$) boson contains the complete information about a state of polarisation of the $Z$ ($W$) boson, and $HZZ$, $HZ\gamma$ and $HWW$ interactions may be studied in detail from a careful analysis of these matrices. 
In this paper, a systematic approach to analyse these polarisation density matrices is presented. 
With the aim of making maximum use of the polarisation information, all of the elements of the polarisation density matrices are related with observables, which are measurable at the environment of $pp$ collisions. 
Consequences of non-standard $HZZ$, $HZ\gamma$ and $HWW$ interactions for these observables are discussed.

\end{abstract}




\newpage

\setcounter{footnote}{0}

\tableofcontents


\section{Introduction}\label{sec:intro}

The ATLAS and CMS collaborations have observed the Higgs boson~\cite{Aad:2012tfa, Chatrchyan:2012xdj}. 
The measurements of the Higgs boson couplings to the standard model (SM) particles are essential tests of the SM. The Higgs boson couplings to the weak bosons, $HZZ$ and $HWW$, can be measured by studying the production cross sections times branching fractions~\cite{Heinemeyer:2013tqa}, 
by studying the angular distribution in the decay processes $H\to ZZ^*_{} \to 4l$ and $H\to WW^*_{} \to l\nu l\nu$~\cite{DellAquila:1985mtb, Nelson:1986ki, Barger:1993wt, Kramer:1993jn, Choi:2002jk, Buszello:2002uu, Hagiwara:2009wt, Gao:2010qx, Bolognesi:2012mm}
and by studying kinematics of the production processes for instance the jet azimuthal angle correlation in the weak boson fusion production~\cite{Plehn:2001nj, Hankele:2006ma, Hagiwara:2009wt, Englert:2012xt, Nakamura:2016agl}. 
The ATLAS and CMS collaborations have already performed measurements of $HZZ$ and $HWW$ interactions by using the above three approaches 
in refs.~\cite{Aad:2013wqa, Aad:2014eva, Aad:2015gba, Aad:2015ona, Khachatryan:2016vau, Chatrchyan:2013iaa, Khachatryan:2014ira, Khachatryan:2014jba}, 
in refs.~\cite{Aad:2013xqa, Aad:2015rwa, Aad:2015mxa, Chatrchyan:2012jja, Chatrchyan:2013mxa, Chatrchyan:2013iaa, Khachatryan:2014kca} and 
in refs.~\cite{Aad:2015tna, Aad:2016nal, Khachatryan:2016tnr}, respectively. 
The production processes $pp \to ZH$ and $pp \to W^{\pm}_{}H$~\cite{Glashow:1978ab, Barger:1986jt, Kniehl:1990iva, HAN1991167, Ohnemus:1992bd, Ciccolini:2003jy, Brein:2003wg, Ferrera:2011bk,Brein:2011vx, Denner:2011id, Dawson:2012gs, Altenkamp:2012sx, Brein:2012ne} provide direct access to the $HZZ$ and $HWW$ couplings, respectively. In these processes, not only the three approaches mentioned above, but also decay properties of the $Z$ and $W$ bosons can be used to study the $HZZ$ and $HWW$ interactions~\cite{Christensen:2010pf, Desai:2011yj, Godbole:2013lna, Delaunay:2013npa, Maltoni:2013sma, Anderson:2013afp, Godbole:2014cfa, Dwivedi:2016xwm, Ferreira:2016jea, Alioli:2017ces}. \\

An arbitrary polarised state of a massive particle with spin $s$ is entirely described by a $(2s+1) \times (2s+1)$ polarisation density matrix; see e.g. refs.~\cite{landau:1965, Schiff:2015, BOURRELY198095, Craigie:1984tk}.
When a particle with spin 1 is in an either completely or partially polarised state, it has 9 independent angular distributions of its decay products at its rest frame and the coefficients of these distributions are written in terms of all the elements of the polarisation density matrix; see e.g. refs.~\cite{BOURRELY198095, Craigie:1984tk}. An experimental determination of the complete decay angular distributions of a particle's decay products is, therefore, identical to an attempt to determine the polarisation density matrix, from which detailed information in reactions which produce that particle can be extracted; see e.g. ref.~\cite{Hagiwara:1986vm} for a theoretical study and e.g. refs.~\cite{Abbiendi:2000ei, Abbiendi:2003wv, Abdallah:2008sf} for experimental studies~\footnote{A general polarisation density matrix $\rho$ for a spin 1 massive particle has 8 degrees of freedom with the normalisation condition $tr(\rho)=1$, and it is possible to parametrise it with 8 real parameters; see e.g. refs.~\cite{BOURRELY198095, Craigie:1984tk}. Recent studies which relate these 8 parameters with observables can be found in refs.~\cite{Ots:2004hk,Ots:2006dv,Boudjema:2009fz,Aguilar-Saavedra:2015yza,Rahaman:2016pqj,Aguilar-Saavedra:2017zkn,Rahaman:2017qql}}. \\

As the $Z$ boson in the process $e^+_{}e^-_{} \to ZH$ can be in a polarised state~\cite{Kelly:1980ue, Rattazzi:1988ye, Barger:1993wt, Hagiwara:1993sw}, so the $Z$ boson in the process $pp \to ZH$ can be. The $W^+_{}$ and $W^-_{}$ in the process $pp \to W^{\pm}_{}H$ can be also in polarised states. It would be, therefore, possible to study $HZZ$, $HZ\gamma$ and $HWW$ interactions in detail from a careful analysis of these states of polarisation. As we mentioned in the previous paragraph, the polarisation density matrix of the $Z$ ($W$) boson contains the complete information about a state of polarisation of the $Z$ ($W$) boson. Hence all of the elements of the polarisation density matrix should be made use of in such a careful analysis. 
In this paper, we present a systematic approach to analyse the polarisation density matrices of the $Z$ boson and $W^{\pm}_{}$ in the processes $pp \to ZH$ and $pp \to W^{\pm}_{}H$. With the aim of making maximum use of the information about states of polarisation, we relate all of the elements of the polarisation density matrices with observables which we can measure at the environment of proton-proton ($pp$) collisions. We discuss consequences of non-standard $HZZ$, $HZ\gamma$ and $HWW$ interactions for these observables. 
A detailed analysis of the polarisation of the $Z$ boson in the process $e^+_{}e^-_{} \to ZH$ is present in ref.~\cite{Hagiwara:1993sw}. Our approach in the process $pp \to ZH$ is an extension of that work to $pp$ collisions. \\

The paper is organised as follows. 
In Section~\ref{sec:helicityamp}, we introduce all the ingredients needed for our analyses in the following sections. At first, we give non-standard $HZZ$, $HZ\gamma$ and $HWW$ couplings. We present the complete helicity amplitudes for the sub-processes $q\bar{q}\to ZH$, $u\bar{d}\to W^+_{}H$ and $d\bar{u}\to W^-_{}H$. These amplitudes are given for the non-standard couplings. We derive the relations between the helicity amplitudes imposed by the CP and $\mathrm{CP\widetilde{T}}$ symmetries. 
In Section~\ref{sec:densitymatrix}, we analyse the polarisation density matrices of the $Z$ boson in the process $pp \to ZH$ in detail. 
First of all, we define the polarisation density matrices consisting of the helicity amplitudes of the previous section. 
Then we derive the 4 different differential cross sections for the process $pp \to ZH$ followed by $Z \to f\bar{f}$ with respect to the $Z$ decay angles. Among the $36$ coefficients ($=9 \times 4$) of these 4 different differential angular distributions, only $15$ coefficients can be non-zero. These coefficients are written in terms of the elements of the polarisation density matrices. 
The restrictions on these coefficients imposed by the CP and $\mathrm{CP\widetilde{T}}$ symmetries are clarified. 
We focus on the coefficients which are strictly zero in the SM due to CP invariance or small in the SM due to the smallness of re-scattering effects, and study the influences of the non-standard $HZZ$ and $HZ\gamma$ couplings. 
In Section~\ref{sec:polWboson}, we analyse the polarisation density matrices of the $W^{+}_{}$ and $W^{-}_{}$ in the process $pp \to W^{\pm}_{}H$ following the same procedure as the previous section. 
Section~\ref{sec:summary} gives a summary.

\section{Constituents of the polarisation density matrices}\label{sec:helicityamp}

As the base vectors for the density matrices~\footnote{Hereafter we call a polarisation density matrix a density matrix for short.}, we choose the eigenfunctions of the helicity operator. The density matrices are, therefore, constructed of helicity amplitudes. 
In this section we derive the complete helicity amplitudes for the sub-processes $q\bar{q}\to ZH$, $u\bar{d}\to W^+_{}H$ and $d\bar{u}\to W^-_{}H$.

\subsection{Effective Lagrangian with CP-odd operators}

\begin{figure}[t]
\centering
\includegraphics[scale=0.45]{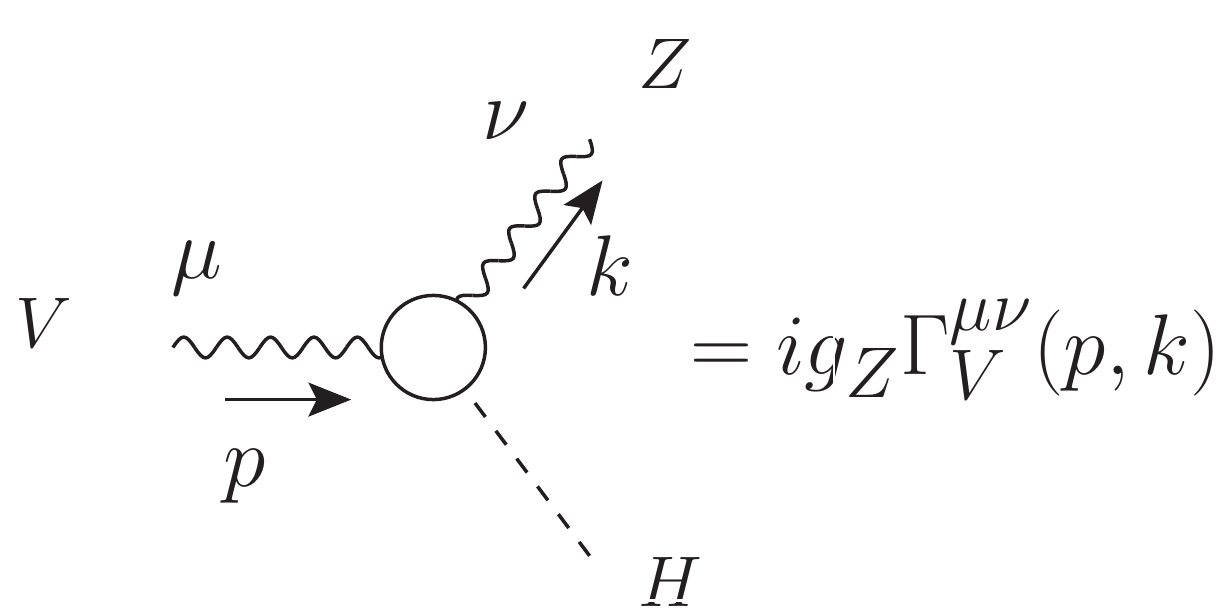}
\hspace{1.5cm}
\includegraphics[scale=0.45]{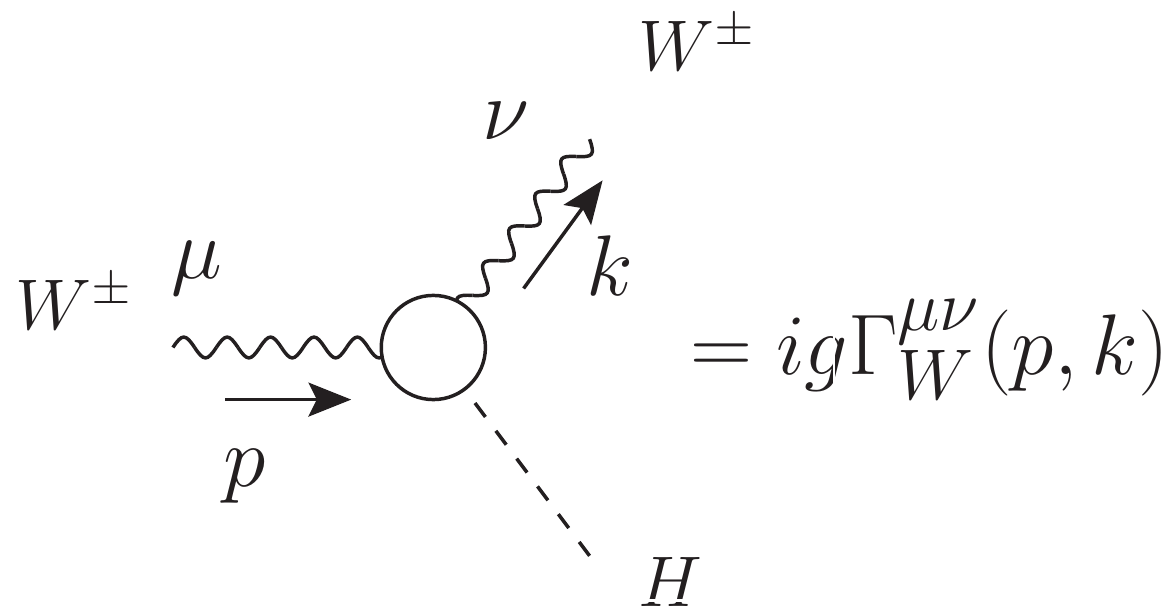}
\caption{\small 
Feynman rules for the $HZV$ ($V=Z,\gamma$) vertex and the $HWW$ vertex. }
\label{figure:vertex}
\end{figure}

We derive non-standard $HZZ$, $HZ\gamma$ and $HWW$ couplings from the following effective Lagrangian~\cite{Hagiwara:1993sw, Hagiwara:2000tk}:
\begin{align}
{\cal L}_{\mathrm{eff}}^{} 
= \bigl( 1 + a_Z^{} ) \frac{1}{2} g_Z^{}m_Z^{} HZ_{\mu}^{}Z^{\mu}_{}
+ \frac{g_Z^{}}{2 m_Z^{}} \sum_{V=Z,\gamma}^{} \Bigl\{
b_V^{} HZ_{\mu\nu}^{} V^{\mu\nu}_{} +
c_V^{} \bigl[ (\partial^{\mu}_{}H) Z^{\nu}_{} - (\partial^{\nu}_{}H) Z^{\mu}_{} \bigr] V_{\mu\nu}^{}  \nonumber \\
+ \frac{1}{2}\widetilde{b}_V^{} \epsilon^{\mu\nu\rho\sigma}_{} H Z_{\mu\nu}^{} V_{\rho\sigma}^{}
\Bigr\} \nonumber \\ 
+ \bigl( 1 + a_W^{} ) g m_W^{} HW^{\dagger}_{\mu}W^{\mu}_{}
+ \frac{g}{m_W^{}} \Bigl\{
b_W^{} HW_{\mu\nu}^{\dagger} W^{\mu\nu}_{} +
c_W^{} \bigl[ (\partial_{\mu}^{}H) W_{\nu}^{\dagger} W^{\mu\nu}_{} - (\partial^{\nu}_{}H) W^{\mu}_{} W_{\mu\nu}^{\dagger} \bigr]  \nonumber \\
+ \frac{1}{2}\widetilde{b}_W^{} \epsilon^{\mu\nu\rho\sigma}_{} H W_{\mu\nu}^{\dagger} W_{\rho\sigma}^{}
\Bigr\},
\label{eq:lagrangian}
\end{align}
where $V_{\mu\nu}^{} = \partial_{\mu}^{} V_{\nu}^{} - \partial_{\nu}^{} V_{\mu}^{}$, $W_{\mu\nu}^{} = \partial_{\mu}^{} W_{\nu}^{} - \partial_{\nu}^{} W_{\mu}^{}$, $W^{\mu}_{}$ is the $W^-_{}$ field, $g=e/\sin{\theta_w^{}}$, $g_Z^{}=g/\cos{\theta_w^{}}$ with $\theta_w^{}$ being the electroweak mixing angle and $e$ being the proton charge, and $\epsilon_{0123}^{}=+1$ in our convention. 
All of the eleven coefficients $a_Z^{}$, $a_W^{}$, $b_V^{}$, $c_V^{}$ and $\widetilde{b}_V^{}$ ($V=Z,\gamma,W$) are real so that ${\cal L}_{\mathrm{eff}}^{} = {\cal L}_{\mathrm{eff}}^{\dagger}$ and zero at the tree level in the SM. The operators whose coefficients are $a_Z^{}$, $a_W^{}$, $b_V^{}$ and $c_V^{}$ ($V=Z,\gamma,W$) are CP-even and the operators whose coefficients are $\widetilde{b}_V^{}$ ($V=Z,\gamma,W$) are CP-odd. The Lagrangian with only the CP-even operators indicates that $H$ is a CP-even scalar field, while the Lagrangian with only the CP-odd operators indicates that $H$ is a CP-odd scalar field. In both cases, the theory is CP invariant. If both the CP-even operator(s) and the CP-odd operator(s) exist, $H$ is no longer a CP eigenstate and the theory is not CP invariant. We call $a_Z^{}$, $a_W^{}$, $b_V^{}$ and $c_V^{}$ ($V=Z,\gamma,W$) CP-even coefficients and $\widetilde{b}_V^{}$ ($V=Z,\gamma,W$) CP-odd coefficients. \\

If we assign the momenta as shown in Figure~\ref{figure:vertex}~\footnote{All pictures in this paper are drawn by using the program JaxoDraw~\cite{Binosi:2003yf}.}, Feynman rules for the $HZV$ ($V=Z,\gamma$) vertex and the $HWW$ vertex can be expressed as
\begin{subequations}
\begin{align}
\Gamma^{\mu \nu}_V (p,k) & =m_Z^{} \biggl( \delta_{ZV}^{} + h_1^V + h_2^V \frac{\hat{s}}{m_Z^2} \biggr) g^{\mu \nu}_{} + h_3^V \frac{k^{\mu}_{} p^{\nu}_{}}{m_Z^{}} +\widetilde{h}_4^V \epsilon^{\mu \nu \alpha \beta}_{} \frac{p_{\alpha}^{}k_{\beta}^{}}{m_Z^{}}, \\
\Gamma^{\mu \nu}_W (p,k) & =m_W^{} \biggl( 1 + h_1^W + h_2^W \frac{\hat{s}}{m_W^2} \biggr) g^{\mu \nu}_{} + h_3^W \frac{k^{\mu}_{} p^{\nu}_{}}{m_W^{}} +\widetilde{h}_4^W \epsilon^{\mu \nu \alpha \beta}_{} \frac{p_{\alpha}^{}k_{\beta}^{}}{m_W^{}},
\end{align}
\end{subequations}
where $\delta_{ZV}^{}=1$ for $V=Z$, $\delta_{ZV}^{}=0$ for $V=\gamma$, and $\hat{s}=(p)^2_{}$. Terms proportional to $k^{\nu}_{}$ vanish for the on-shell $Z$ boson and the on-shell $W$ boson, and terms proportional to $p^{\mu}_{}$ also vanish in our processes where the intermediate off-shell vector boson ($Z$, $\gamma$ or $W$) couples to the four-vector consisting of the two massless quarks. These terms are, therefore, not included in the above formulae. All the form factors are constant and expressed in terms of the coefficients in the effective Lagrangian in eq.~(\ref{eq:lagrangian}):
\begin{align}
h_1^Z & = a_Z^{}+b_Z^{} - (b_Z^{}-c_Z^{}) \frac{m_H^2}{m_Z^2}, 
& h_1^{\gamma} & = \frac{1}{2} (b_{\gamma}^{}-c_{\gamma}^{}) \frac{m_Z^2-m_H^2}{m_Z^2}, 
& h_1^W & = a_W^{}+b_W^{} - (b_W^{}-c_W^{}) \frac{m_H^2}{m_W^2}, \nonumber \\
h_2^Z & = b_Z^{}, 
& h_2^{\gamma} & = \frac{1}{2} ( b_{\gamma}^{} + c_{\gamma}^{} ), 
& h_2^W & = b_W^{}, \nonumber \\
h_3^Z & = - 2(b_Z^{}-c_Z^{}), 
& h_3^{\gamma} & = - (b_{\gamma}^{}-c_{\gamma}^{}), 
& h_3^W & = - 2(b_W^{}-c_W^{}), \nonumber \\
\widetilde{h}_4^Z & = -2 \widetilde{b}_Z^{}, 
& \widetilde{h}_4^{\gamma} & = - \widetilde{b}_{\gamma}^{}, 
& \widetilde{h}_4^W & = -2 \widetilde{b}_W^{}.
\end{align}
The form factors $h_i^V$ $(V=Z, \gamma, W)$ $(i=1,2,3)$ consist of the CP-even coefficients and $\widetilde{h}_4^V$ $(V=Z, \gamma, W)$ consist of the CP-odd coefficients. We call $h_i^V$ $(V=Z, \gamma, W)$ $(i=1,2,3)$ CP-even form factors and $\widetilde{h}_4^V$ $(V=Z, \gamma, W)$ CP-odd form factors.


\begin{figure}[t]
\centering
\includegraphics[scale=0.5]{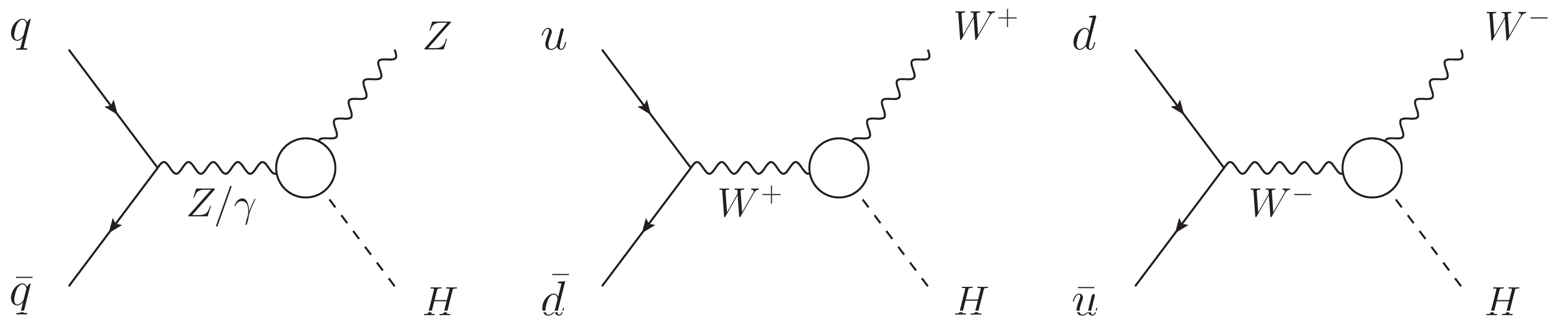}
\caption{\small 
Feynman diagrams for the sub-processes $q\bar{q}\to ZH$, $u\bar{d}\to W^+_{}H$ and $d\bar{u}\to W^-_{}H$. The circles denote the non-standard couplings derived from the effective Lagrangian in eq.~(\ref{eq:lagrangian}).}
\label{figure:diagrams}
\end{figure}

\begin{figure}[t]
\centering
\includegraphics[scale=0.5]{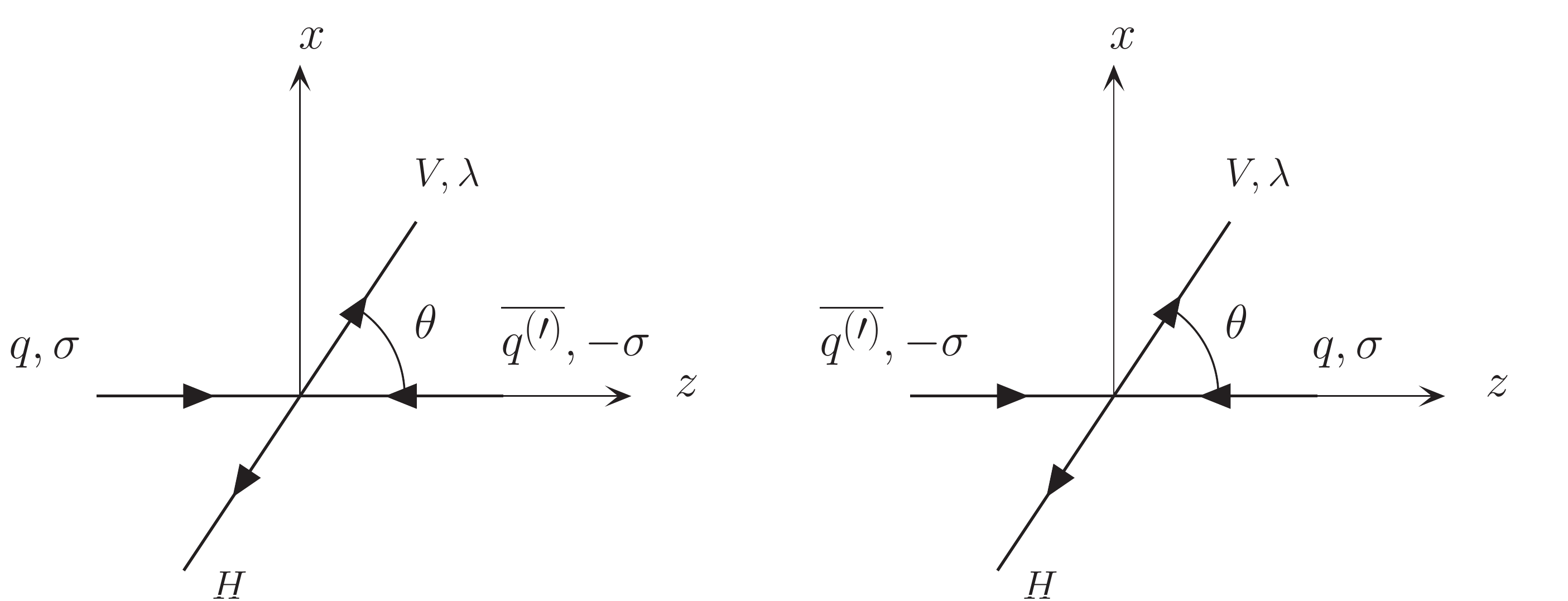}
\caption{\small 
{\it Left}: the $q\overline{q^{(\prime)}_{}}$ c.m. frame, where the quark ($q$) moves along the positive direction of the $z$-axis. The production amplitudes ${\cal M}_{\sigma}^{\lambda}(q \overline{q^{(\prime)}_{}})$ are evaluated in this frame. 
{\it Right}: the $\overline{q^{(\prime)}_{}}q$ c.m. frame, where the antiquark ($\overline{q^{(\prime)}_{}}$) moves along the positive direction of the $z$-axis. The production amplitudes ${\cal M}_{\sigma}^{\lambda}(\overline{q^{(\prime)}_{}}q)$ are evaluated in this frame. The quark helicity $\sigma(=\pm1)$, the antiquark helicity $-\sigma$ and the weak boson helicity $\lambda(=\pm1,0)$ are shown, and the polar angle $\theta$ of the weak boson $V$ from the $z$-axis is denoted in both the frames. 
}
\label{figure:frames}
\end{figure}

\subsection{Helicity amplitudes}\label{sec:helamp}

Feynman diagrams for the sub-processes $q\bar{q}\to ZH$, $u\bar{d}\to W^+_{}H$ and $d\bar{u}\to W^-_{}H$ are shown in Figure~\ref{figure:diagrams}. The circles denote the non-standard couplings derived in the previous section. We evaluate production helicity amplitudes in the following two frames. Let us assume that a direction of the $z$-axis is fixed along $pp$ collisions. In one frame, which we call the $q\overline{q^{(\prime)}_{}}$ centre-of-mass (c.m.) frame and is shown in the left picture of Figure~\ref{figure:frames}, the quark ($q$) moves along the positive direction of the $z$-axis. In another frame, which we call the $\overline{q^{(\prime)}_{}}q$ c.m. frame and is shown in the right picture of Figure~\ref{figure:frames}, the antiquark ($\overline{q^{(\prime)}_{}}$) moves along the positive direction of the $z$-axis. 
In Figure~\ref{figure:frames}, our notation for helicities and the polar angle $\theta$ are also shown. 
We neglect the masses of $q$ and $\overline{q^{(\prime)}_{}}$, thus the helicity of $\overline{q^{(\prime)}_{}}$ is always opposite to that of $q$ in our sub-processes. Helicity amplitudes are, therefore, given for the quark helicity $\sigma(=\pm1)$ and the weak boson helicity $\lambda(=\pm1,0)$; the antiquark helicity is automatically fixed to $-\sigma$~\footnote{The fermion helicity is always normalised to $\pm1$ in this paper.}. 
We denote the helicity amplitudes evaluated in the $q\overline{q^{(\prime)}_{}}$ c.m. frame by ${\cal M}_{\sigma}^{\lambda}(q \overline{q^{(\prime)}_{}})$ and those evaluated in the $\overline{q^{(\prime)}_{}}q$ c.m. frame by ${\cal M}_{\sigma}^{\lambda}(\overline{q^{(\prime)}_{}}q)$. 
It should be emphasised that $\sigma$ always denotes the helicity of the quark and $\theta$ is the polar angle of the weak boson from the $z$-axis (not the quark momentum direction). 
The helicity amplitudes ${\cal M}_{\sigma}^{\lambda}(q \bar{q})$ for the sub-process $q\bar{q}\to ZH$ are
\begin{subequations}\label{eq:ZampAll1}
\begin{align}
{\cal M}_{\sigma}^{\lambda=\pm}(q \bar{q}) &= \sigma \frac{1+\sigma \lambda \cos{\theta}}{\sqrt{2}} \hat{M}_{\sigma}^{\lambda=\pm}, \label{eq:Zamp1} \\
{\cal M}_{\sigma}^{\lambda=0}(q \bar{q}) &= \sin{\theta} \hat{M}_{\sigma}^{\lambda=0}, \label{eq:Zamp2}
\end{align}
\end{subequations}
where
\begin{align}
\hat{M}_{\sigma}^{\lambda=\pm} & =
g_Z^{} m_Z^{} \sqrt{\hat{s}} \Biggl[
\frac{g_{Z \sigma}^{}}{\hat{s} - m_Z^2} \biggl( 1+ h_1^Z + h_2^Z \frac{\hat{s}}{m_Z^2} + i \lambda \widetilde{h}_4^Z \frac{k\sqrt{\hat{s}}}{m_Z^2} \biggr) +
\frac{Q_q^{}e}{\hat{s}} \biggl( h_1^{\gamma} + h_2^{\gamma} \frac{\hat{s}}{m_Z^2} + i \lambda \widetilde{h}_4^{\gamma} \frac{k\sqrt{\hat{s}}}{m_Z^2} \biggr) \Biggr], \nonumber \\
\hat{M}_{\sigma}^{\lambda=0} & = -
g_Z^{} w \sqrt{\hat{s}} \Biggl[
\frac{g_{Z \sigma}^{}}{\hat{s} - m_Z^2} \biggl( 1+h_1^Z + h_2^Z \frac{\hat{s}}{m_Z^2} + h_3^Z \frac{k^2_{}\sqrt{\hat{s}}}{m_Z^2w} \biggr) +
\frac{Q_q^{}e}{\hat{s}} \biggl( h_1^{\gamma} + h_2^{\gamma} \frac{\hat{s}}{m_Z^2} + h_3^{\gamma} \frac{k^2_{}\sqrt{\hat{s}}}{m_Z^2w}  \biggr) \Biggr].
\label{eq:Zsubamplitudes}
\end{align} 
Here $\sqrt{\hat{s}}$ is the $q\bar{q}$ c.m. energy, $w$ is the energy of the $Z$ boson: $w = (\hat{s}+m_Z^2-m_H^2)/(2\sqrt{\hat{s}})$, $k$ is the momentum of the $Z$ boson: $k=\sqrt{w^2_{}-m_Z^2}$, $Q_q^{}$ is the electric charge of the quark in units of $e$, $g_{Z+}^{} = g_Z^{} ( - Q_q^{} \sin^2_{}{\theta_w^{}})$, and $g_{Z-}^{} = g_Z^{} (T_q^3 - Q_q^{} \sin^2_{}{\theta_w^{}})$ where $T_u^3=1/2$ and $T_d^3=-1/2$. \\

Although the actual calculation of the helicity amplitudes ${\cal M}_{\sigma}^{\lambda}(\bar{q}q)$ is easy, it is also possible to estimate them from ${\cal M}_{\sigma}^{\lambda}(q\bar{q})$ as follows. 
If we denote the helicity of $\bar{q}$ by $\sigma$ and the helicity of $q$ by $-\sigma$ (this is opposite to our helicity notation), the angular part is the same as ${\cal M}_{\sigma}^{\lambda}(q\bar{q})$ and the amplitudes are given by
\begin{subequations}\label{eq:ZampAll2-temp}
\begin{align}
{\cal M}_{}^{\lambda=\pm}\bigl(\bar{q}(\sigma)q(-\sigma) \bigr) &= \sigma \frac{1+\sigma \lambda \cos{\theta}}{\sqrt{2}} \hat{M}_{-\sigma}^{\lambda=\pm}, \label{eq:Zamp3-temp} \\
{\cal M}_{}^{\lambda=0}\bigl(\bar{q}(\sigma)q(-\sigma) \bigr) &= \sin{\theta} \hat{M}_{-\sigma}^{\lambda=0}, \label{eq:Zamp4-temp}
\end{align}
\end{subequations}
where the helicity of each particle is explicitly shown in parenthesis. Because the helicity of a quark is equal to its chirality and the helicity of an antiquark is opposite to its chirality in the massless limit, $q$ with helicity $-\sigma$ has the chirality $-\sigma$. 
Hence the coupling $g_{Z \sigma}^{}$ must be replaced by $g_{Z -\sigma}^{}$ in eq.~(\ref{eq:Zsubamplitudes}), and this replacement is expressed by $\hat{M}_{-\sigma}^{\lambda}$ in eq.~(\ref{eq:ZampAll2-temp}). By the simple replacement $\sigma \to -\sigma$ in eq.~(\ref{eq:ZampAll2-temp}), we obtain the helicity amplitudes ${\cal M}_{\sigma}^{\lambda}(\bar{q}q)$ in our helicity notation:
\begin{subequations}\label{eq:ZampAll2}
\begin{align}
{\cal M}_{\sigma}^{\lambda=\pm}(\bar{q}q) &= -\sigma \frac{1-\sigma \lambda \cos{\theta}}{\sqrt{2}} \hat{M}_{\sigma}^{\lambda=\pm}, \label{eq:Zamp3}\\
{\cal M}_{\sigma}^{\lambda=0}(\bar{q}q) &= \sin{\theta} \hat{M}_{\sigma}^{\lambda=0}. \label{eq:Zamp4} 
\end{align}
\end{subequations}
We note that the relative sign among the amplitudes in eq.~(\ref{eq:ZampAll1}) and that among the amplitudes in eq.~(\ref{eq:ZampAll2}) are important, because they appear in the off-diagonal elements of the density matrices. \\

Similarly, the helicity amplitudes for the sub-process $u\bar{d}\to W^+_{}H$ are given by
\begin{subequations}\label{eq:WpampAll}
\begin{align}
{\cal M}_{-}^{\lambda=\pm}(u \bar{d}) &= - \frac{1 - \lambda \cos{\theta}}{\sqrt{2}} (V^{}_{ud})^*_{}\ \hat{N}_{}^{\lambda=\pm}, \label{eq:Wpamp1} \\
{\cal M}_{-}^{\lambda=0}(u \bar{d}) &= \sin{\theta}\ (V^{}_{ud})^*_{}\ \hat{N}_{}^{\lambda=0}, \label{eq:Wpamp2} \\
{\cal M}_{-}^{\lambda=\pm}(\bar{d}u) &=  \frac{1 + \lambda \cos{\theta}}{\sqrt{2}} (V^{}_{ud})^*_{}\ \hat{N}_{}^{\lambda=\pm}, \label{eq:Wpamp3}\\
{\cal M}_{-}^{\lambda=0}(\bar{d}u) &= \sin{\theta}\ (V^{}_{ud})^*_{}\ \hat{N}_{}^{\lambda=0}, \label{eq:Wpmp4} 
\end{align}
\end{subequations}
and those for the sub-process $d\bar{u}\to W^-_{}H$ are given by
\begin{subequations}\label{eq:WmampAll}
\begin{align}
{\cal M}_{-}^{\lambda=\pm}(d \bar{u}) &= - \frac{1 - \lambda \cos{\theta}}{\sqrt{2}} V^{}_{ud}\ \hat{N}_{}^{\lambda=\pm}, \label{eq:Wpamm1} \\
{\cal M}_{-}^{\lambda=0}(d \bar{u}) &= \sin{\theta}\ V^{}_{ud}\ \hat{N}_{}^{\lambda=0}, \label{eq:Wpamm2} \\
{\cal M}_{-}^{\lambda=\pm}(\bar{u}d) &=  \frac{1 + \lambda \cos{\theta}}{\sqrt{2}} V^{}_{ud}\ \hat{N}_{}^{\lambda=\pm}, \label{eq:Wpamm3}\\
{\cal M}_{-}^{\lambda=0}(\bar{u}d) &= \sin{\theta}\ V^{}_{ud}\ \hat{N}_{}^{\lambda=0}, \label{eq:Wpmm4} 
\end{align}
\end{subequations}
where
\begin{align}
\hat{N}_{}^{\lambda=\pm} & =
\frac{1}{\sqrt{2}}g^2_{} m_W^{} \sqrt{\hat{s}} 
\frac{1}{\hat{s} - m_W^2} \biggl( 1 + h_1^W + h_2^W \frac{\hat{s}}{m_W^2} + i \lambda \widetilde{h}_4^W \frac{k\sqrt{\hat{s}}}{m_W^2} \biggr), \nonumber \\
\hat{N}_{}^{\lambda=0} & = -
\frac{1}{\sqrt{2}}g^2_{} w \sqrt{\hat{s}} 
\frac{1}{\hat{s} - m_W^2} \biggl( 1 + h_1^W + h_2^W \frac{\hat{s}}{m_W^2} + h_3^W \frac{k^2_{}\sqrt{\hat{s}}}{m_W^2 w} \biggr).
\label{eq:Wsubamplitudes}
\end{align} 
Here $w$ is the energy of the $W$ boson: $w = (\hat{s}+m_W^2-m_H^2)/(2\sqrt{\hat{s}})$, $k$ is the momentum of the $W$ boson: $k=\sqrt{w^2_{}-m_W^2}$, and $V_{ud}^{}$ is the element of the Cabibbo-Kobayashi-Maskawa (CKM) matrix. 
\\

By looking at the helicity amplitudes, we can already discuss a difference between the $Z$ boson and the $W$ boson with regard to states of polarisation. 
Since the $W$ boson couples to only a fermion with chirality $-1$, the initial quark $u$ or $d$ always has the helicity $\sigma=-1$. 
Therefore, while the $Z$ boson is in a partially polarised state (often called a mixed state; see e.g. refs.~\cite{landau:1965, Schiff:2015, BOURRELY198095, Craigie:1984tk}), the $W$ boson (both the $W^+_{}$ and $W^-_{}$) is in a completely polarised state (often called a pure state).

\subsection{Requests from symmetries}

\begin{figure}[t]
\centering
\includegraphics[scale=0.42]{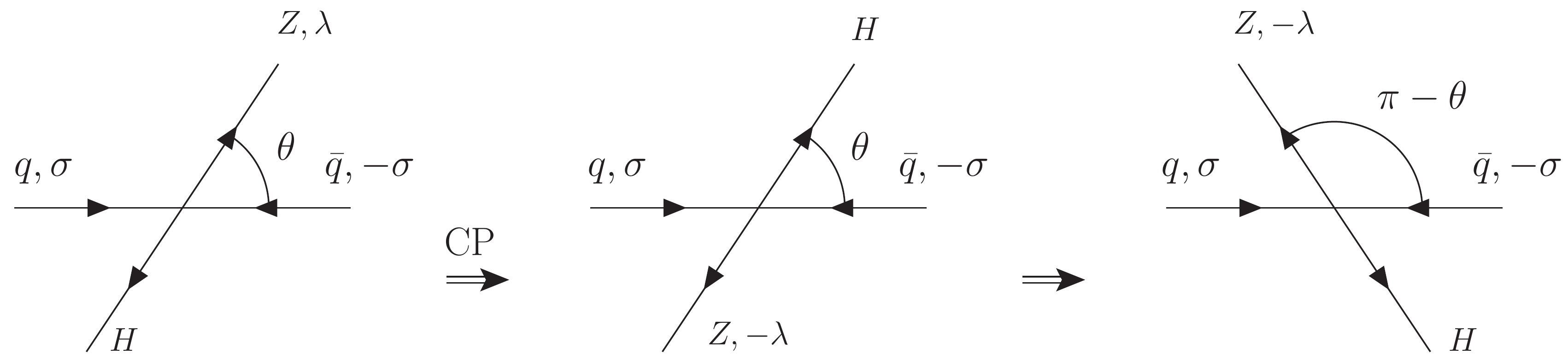}
\includegraphics[scale=0.42]{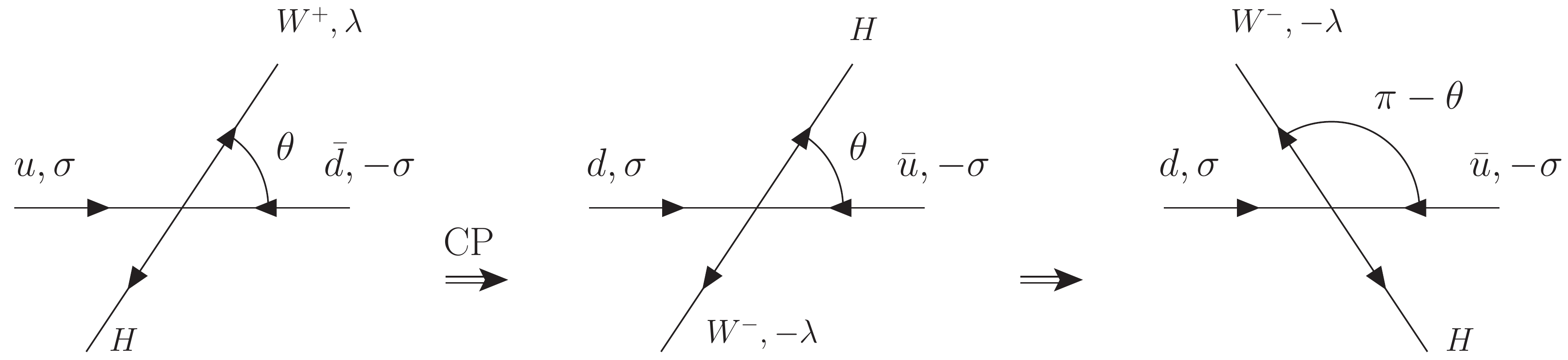}
\caption{\small The original states, the states after CP transformation and the states after the rotation around the $z$-axis by $\pi$ are shown for the process $q\bar{q} \to ZH$ (upper figures) and the process $u\bar{d} \to W^+_{}H$ (lower figures).}
\label{figure:CP}
\end{figure}

Conditions imposed by symmetries lead to certain relations between the helicity amplitudes. The upper figures in Figure~\ref{figure:CP} show the original states in the process $q\bar{q} \to ZH$ (left), the states after CP transformation (middle) and the states after the rotation around the $z$-axis by $\pi$ (right).
From these figures, we find that the invariance under CP leads to the relation 
\begin{align}
{\cal M}_{\sigma}^{\lambda}(q \bar{q}) (\theta) = {\cal M}_{\sigma}^{-\lambda}(q \bar{q}) (\pi-\theta), \label{eq:CPinvZ} 
\end{align}
where ${\cal M}_{\sigma}^{\lambda}(q \bar{q}) (\theta) = {\cal M}_{\sigma}^{\lambda}(q \bar{q})$ given in eq.~(\ref{eq:ZampAll1}), and ${\cal M}_{\sigma}^{-\lambda}(q \bar{q})  (\pi-\theta)$ is obtained by setting $\theta \to \pi - \theta$ in ${\cal M}_{\sigma}^{-\lambda}(q \bar{q})$. 
It is easy to see that a non-zero value of the CP-odd form factors $\widetilde{h}_4^Z$ and/or $\widetilde{h}_4^{\gamma}$ violates this relation. 
Similarly, the lower figures in Figure~\ref{figure:CP} show the original states in the process $u\bar{d} \to W^+_{}H$ (left), the states after CP transformation (middle) and the states after the rotation around the $z$-axis by $\pi$ (right). From these figures, we find that CP invariance requests the relation
\begin{align}
{\cal M}_{-}^{\lambda}(u \bar{d}) (\theta) = {\cal M}_{-}^{-\lambda}(d \bar{u}) (\pi-\theta), \label{eq:CPinvW}
\end{align}
where ${\cal M}_{-}^{\lambda}(u \bar{d}) (\theta) = {\cal M}_{-}^{\lambda}(u \bar{d})$ given in eq.~(\ref{eq:WpampAll}), and ${\cal M}_{-}^{-\lambda}(d \bar{u}) (\pi-\theta)$ is obtained by setting $\theta \to \pi - \theta$ in ${\cal M}_{-}^{-\lambda}(d \bar{u})$ given in eq.~(\ref{eq:WmampAll}). This relation is violated by the imaginary part of the element $V_{ud}^{}$ of the CKM matrix, even when the CP-odd form factor $\widetilde{h}_4^W$ is zero. 
However, this CP violation phase in $V_{ud}^{}$~\cite{Kobayashi:1973fv} is always regarded as an overall common phase among the amplitudes. 
For instance, the phase in $V^{}_{ud}$ is an overall common phase among ${\cal M}_{-}^{\lambda=+}(d \bar{u})$, ${\cal M}_{-}^{\lambda=-}(d \bar{u})$ and ${\cal M}_{-}^{\lambda=0}(d \bar{u})$. 
Therefore, the effect of the CP violation phase in the CKM matrix never appears even in the off-diagonal elements of the density matrices. Only a non-zero value of the CP-odd form factor $\widetilde{h}_4^W$ violates the relation in eq.~(\ref{eq:CPinvW}) {\it and} affects a state of polarisation of the $W$ boson. 
We can write the relations in eqs.~(\ref{eq:CPinvZ}) and (\ref{eq:CPinvW}) in the following convenient forms, respectively:
\begin{subequations}
\begin{align}
\hat{M}_{\sigma}^{\lambda} & = \hat{M}_{\sigma}^{-\lambda}, \label{eq:CPinvZ-2}\\
\hat{N}_{}^{\lambda} & = \hat{N}_{}^{-\lambda}. 
\end{align}
\end{subequations}

CPT invariance {\it and} the unitarity condition conclude~\cite{Hagiwara:1986vm} that the following relations hold at the tree level approximation and violation of the relations indicates the existence of re-scattering effects:
\begin{subequations}\label{eq:CPTinv}
\begin{align}
{\cal M}_{\sigma}^{\lambda}(q \bar{q}) (\theta) &= \bigl\{ {\cal M}_{\sigma}^{-\lambda}(q \bar{q}) (\pi-\theta) \bigr\}^*_{}, \label{eq:CPTinvZ} \\
{\cal M}_{-}^{\lambda}(u \bar{d}) (\theta) &= \bigl\{ {\cal M}_{-}^{-\lambda}(d \bar{u}) (\pi-\theta) \bigr\}^*_{}. \label{eq:CPTinvW}
\end{align}
\end{subequations}
It is easy to see that the relations hold if all of the form factors are real. 
Following ref.~\cite{Hagiwara:1986vm}, we call the invariance which leads to these relations $\mathrm{CP\widetilde{T}}$ invariance. 
We can write the relations in eq.~(\ref{eq:CPTinv}) in the following convenient forms, respectively:
\begin{subequations}
\begin{align}
\hat{M}_{\sigma}^{\lambda} & = \bigl( \hat{M}_{\sigma}^{-\lambda} \bigr)^*_{}, \label{eq:CPTinvZ-2}\\
\hat{N}_{}^{\lambda} & = \bigl( \hat{N}_{}^{-\lambda} \bigr)^*_{}. 
\end{align}
\end{subequations}

\section{Polarisation of the $Z$ boson}\label{sec:densitymatrix}

\subsection{Polarisation density matrices of the $Z$ boson}\label{sec:derivation}

In this section we define the density matrices of the $Z$ boson by using our notation of the helicity amplitudes of the previous section. We consider the sub-process
\begin{align}
q(\sigma) + \bar{q}( -\sigma ) & \to Z(\lambda) + H; \nonumber \\
Z(\lambda) & \to f(\tau) + \bar{f}(-\tau),
\end{align}
where the helicity of each particle is shown in parenthesis. We neglect the masses of the final fermion $f$ and the final antifermion $\bar{f}$, thus the helicity of $\bar{f}$ is always opposite to that of $f$. We express the full helicity amplitude as
\begin{align}
{\cal T}_{\sigma}^{\tau}(q\bar{q}) =  P_Z^{} \sum_{\lambda=\pm,0}^{} {\cal M}_{\sigma}^{\lambda}(q \bar{q})\ D_{\lambda}^{\tau},\label{eq:fullampZ}
\end{align}
where the production amplitude ${\cal M}_{\sigma}^{\lambda}(q \bar{q})$ is given in eq.~(\ref{eq:ZampAll1}), $D_{\lambda}^{\tau}$ is the decay helicity amplitude, and
\begin{align}
P_Z^{} =  ( Q^2_{} - m_Z^2 + i m_Z^{} \Gamma_Z^{} )^{-1}_{}
\end{align}
denotes the propagator factor of the $Z$ boson. 
We evaluate the decay amplitude in the following four-momentum frame:
\begin{align}
Z:&\ \ \bigl( m_Z^{}, 0, 0, 0 \bigr) \nonumber \\
f:&\ \  \frac{m_Z^{}}{2}\bigl( 1,\ \sin{\widehat{\theta}} \cos{\widehat{\phi}},\ \sin{\widehat{\theta}} \sin{\widehat{\phi}},\ \cos{\widehat{\theta}} \bigr) \nonumber \\
\bar{f} :&\ \  \frac{m_Z^{}}{2}\bigl( 1,\ -\sin{\widehat{\theta}} \cos{\widehat{\phi}},\ -\sin{\widehat{\theta}} \sin{\widehat{\phi}},\ -\cos{\widehat{\theta}} \bigr). \label{eq:decaykinematics}
\end{align}
The decay amplitude is
\begin{align}
D_{\lambda}^{\tau} = g_{Zf\bar{f}}^{\tau}\ m_Z^{}\ d_{\lambda}^{\tau},
\end{align}
where
\begin{subequations}\label{eq:decayamp}
\begin{align}
d_{\lambda=\pm}^{\tau} & = \tau \frac{1+\lambda \tau \cos{\widehat{\theta}}}{\sqrt{2}} e^{i \lambda \widehat{\phi}}_{}, \\
d_{\lambda=0}^{\tau} & = \sin{\widehat{\theta}}.
\end{align}
\end{subequations}
The explicit forms of the coupling $g_{Zf\bar{f}}^{\tau}$ are, $g_{Zl\bar{l}}^{+} = g_Z^{} \sin^2_{}{\theta_w^{}}$ and $g_{Zl\bar{l}}^{-} = g_Z^{} ( -1/2+\sin^2_{}{\theta_w^{}} )$ for a charged lepton pair, $g_{Zu\bar{u}}^{+} =  -(2/3)g_Z^{}\sin^2_{}{\theta_w^{}}$ and $g_{Zu\bar{u}}^{-} = g_Z^{} (1/2-(2/3)\sin^2_{}{\theta_w^{}})$ for a up-type quark pair, $g_{Zd\bar{d}}^{+} = (1/3)g_Z^{} \sin^2_{}{\theta_w^{}}$ and $g_{Zd\bar{d}}^{-} = g_Z^{} (-1/2+(1/3)\sin^2_{}{\theta_w^{}})$ for a down-type quark pair. A straightforward manipulation gives 
\begin{align}
\sum_{\sigma}^{} \bigl| {\cal T}_{\sigma}^{\tau}(q\bar{q}) \bigr|^2_{}
 & = \bigl|P_Z^{} m_Z^{} g_{Zf\bar{f}}^{\tau} \bigr|^2_{}
\sum_{\sigma}^{}\sum_{\lambda^{\prime}_{}, \lambda} \bigl( d_{\lambda^{\prime}}^{\tau} \bigr)^*_{} \rho_{\sigma}^{\lambda^{\prime} \lambda}(q\bar{q}) d_{\lambda}^{\tau} \nonumber \\
 & = \bigl|P_Z^{} m_Z^{} g_{Zf\bar{f}}^{\tau} \bigr|^2_{}
 d_{}^{\tau \dagger} \sum_{\sigma}^{} \rho_{\sigma}^{}(q\bar{q}) d_{}^{\tau}, \label{eq:ampsquaredZ}
\end{align}
where
\begin{align}
\sum_{\sigma} \rho_{\sigma}^{\lambda^{\prime} \lambda}(q\bar{q}) = \sum_{\sigma}
 \bigl\{ {\cal M}_{\sigma}^{\lambda^{\prime}_{}}(q \bar{q}) \bigr\}^{\ast}_{}
 {\cal M}_{\sigma}^{\lambda}(q \bar{q})\label{eq:densityMqqbarZ}
\end{align}
represents the elements of the density matrix in the helicity basis of the $Z$ boson in the $q\bar{q}$ c.m. frame, and at the last equality the following $3\times3$ matrix form is employed:
\begin{align}
\rho_{\sigma}^{}(q\bar{q}) = 
\left(
\begin{array}{ccc}
\rho_{\sigma}^{++}(q\bar{q}) & \rho_{\sigma}^{+-}(q\bar{q}) & \rho_{\sigma}^{+0}(q\bar{q}) \\
\rho_{\sigma}^{-+}(q\bar{q}) & \rho_{\sigma}^{--}(q\bar{q}) & \rho_{\sigma}^{-0}(q\bar{q}) \\
\rho_{\sigma}^{0+}(q\bar{q}) & \rho_{\sigma}^{0-}(q\bar{q}) & \rho_{\sigma}^{00}(q\bar{q}) \\
\end{array}
\right)
,\ \ \ 
d^{\tau}_{} = 
\left(
\begin{array}{c}
d^{\tau}_{+} \\
d^{\tau}_{-} \\
d^{\tau}_{0} \\
\end{array}
\right).\label{eq:matrixform}
\end{align}
The density matrix is a $3\times3$ Hermitian matrix: $\sum_{\sigma} \rho_{\sigma}^{}(q\bar{q}) = ( \sum_{\sigma} \rho_{\sigma}^{}(q\bar{q}) )^{\dagger}_{}$~\footnote{In general a density matrix $\rho$ has a normalisation condition $tr(\rho)=1$; see e.g. refs.~\cite{landau:1965, Schiff:2015, BOURRELY198095, Craigie:1984tk}. However, we employ the non-normalised form such as eq.~(\ref{eq:densityMqqbarZ}) and call it a density matrix in this paper. The degrees of freedom of our density matrices is, therefore, not 8 but 9.}.\\

The full helicity amplitude ${\cal T}_{\sigma}^{\tau}(\bar{q}q)$, in which the production amplitude is evaluated in the $\bar{q}q$ c.m. frame shown in the right picture of Figure~\ref{figure:frames}, can be treated in the same manner:
\begin{subequations}
\begin{align}
{\cal T}_{\sigma}^{\tau}(\bar{q}q) & =  P_Z^{} \sum_{\lambda=\pm,0}^{} {\cal M}_{\sigma}^{\lambda}(\bar{q}q)\ D_{\lambda}^{\tau}, \label{eq:fullampZqbarq} \\
\sum_{\sigma}^{} \bigl| {\cal T}_{\sigma}^{\tau}(\bar{q}q) \bigr|^2_{}
 & = \bigl|P_Z^{} m_Z^{} g_{Zf\bar{f}}^{\tau} \bigr|^2_{}
\sum_{\sigma}^{}\sum_{\lambda^{\prime}_{}, \lambda} \bigl( d_{\lambda^{\prime}}^{\tau} \bigr)^*_{} \rho_{\sigma}^{\lambda^{\prime} \lambda}(\bar{q}q) d_{\lambda}^{\tau} \nonumber \\
& = \bigl|P_Z^{} m_Z^{} g_{Zf\bar{f}}^{\tau} \bigr|^2_{}
 d_{}^{\tau \dagger} \sum_{\sigma}^{} \rho_{\sigma}^{}(\bar{q}q) d_{}^{\tau}, \label{eq:ampsquaredZqbarq}
\end{align}
\end{subequations}
where 
\begin{align}
\sum_{\sigma} \rho_{\sigma}^{\lambda^{\prime} \lambda}(\bar{q} q) = \sum_{\sigma}
 \bigl\{ {\cal M}_{\sigma}^{\lambda^{\prime}_{}}(\bar{q} q) \bigr\}^{\ast}_{}
 {\cal M}_{\sigma}^{\lambda}(\bar{q} q)\label{eq:densityMqbarqZ}
\end{align}
represents the elements of the density matrix in the helicity basis of the $Z$ boson in the $\bar{q}q$ c.m. frame. The production amplitude ${\cal M}_{\sigma}^{\lambda}(\bar{q} q)$ is given in eq.~(\ref{eq:ZampAll2}).

\subsection{Decay angular distributions of the polarised $Z$ boson}\label{sec:crossZboson}

In terms of the density matrices defined in eqs.~(\ref{eq:densityMqqbarZ}) and (\ref{eq:densityMqbarqZ}), 
the complete differential cross section for the process $pp\to ZH$ followed by $Z\to f\bar{f}$ in the narrow width approximation can be expressed as follows:
\begin{align}
\frac{d\sigma}{d\hat{s}\ dy\ d\cos{\theta}\ d\cos{\widehat{\theta}}\ d\widehat{\phi} }
= & \frac{m_Z^{}k}{ 12288 \pi^3_{} \Gamma_Z^{} s \hat{s}^{\frac{3}{2}}_{}} 
\sum_f \sum_{\tau} \bigl| g_{Zf\bar{f}}^{\tau} \bigr|^2_{} C_f^{} \nonumber \\
& \times \sum_{q} \biggl[ q(x_1^{}) \bar{q}(x_2^{})\ d_{}^{\tau \dagger} \sum_{\sigma}^{} \rho_{\sigma}^{}(q\bar{q}) d_{}^{\tau}
+
\bar{q}(x_1^{}) q(x_2^{})\ d_{}^{\tau \dagger} \sum_{\sigma}^{} \rho_{\sigma}^{}(\bar{q}q) d_{}^{\tau} \biggr],\label{eq:differentialZH}
\end{align}
where $s$ is the c.m. energy squared of the $pp$ collisions, $\hat{s}$ is the c.m. energy squared of the $q\bar{q}$ collisions, 
$y$ is the rapidity of the $q\bar{q}$ c.m. frame and the $\bar{q}q$ c.m. frame (see Figure~\ref{figure:frames}) on the $pp$ c.m. frame (i.e. the experimental frame), 
$C_f^{}$ is an effective colour factor: $C_f^{}=1$ for $Z$ decays into a charged lepton pair and $C_f^{}=3$ for $Z$ decays into a quark pair, $q(x_i^{})$ and $\bar{q}(x_i^{})$ are the quark and antiquark parton distribution functions (PDFs) with energy fraction $x_i^{}$. Summations are performed over the final fermion flavor $f$, the helicity $\tau$ of the final fermion, and the initial quark and antiquark flavor $q$. Averages are performed over the helicity and colour of the initial quark, and those of the initial antiquark. Recall that $k$ is the momentum of the $Z$ boson and $\theta$ is the polar angle of the $Z$ boson in the $q\bar{q}$ c.m. frame and the $\bar{q}q$ c.m. frame. The decay angles $\widehat{\theta}$ and $\widehat{\phi}$ are defined in eq.~(\ref{eq:decaykinematics}). 
Here $x_1^{}$ ($x_2^{}$) denotes the energy fraction of the initial quark or antiquark in the proton that moves along the positive (negative) direction of the $z$-axis, once a direction of the $z$-axis is fixed along the $pp$ collisions. 
The integration variables $\hat{s}$ and $y$ are related to $x_1^{}$ and $x_2^{}$:
\begin{align}
\hat{s} = s x_1^{} x_2^{},\ \ \ y = \frac{1}{2}\ln{\frac{x_1^{}}{x_2^{}}}.
\end{align}
The allowed region is
\begin{align}
(m_Z^{} + m_H^{})^2_{} < \hat{s} < s, \ \ \ - \ln{\sqrt{\frac{s}{\hat{s}}}} < y < \ln{\sqrt{\frac{s}{\hat{s}}}}.\label{eq:allowedregion}
\end{align} 
\\

\renewcommand{\arraystretch}{2.5}

\begin{table}[t]
\centering
\begin{tabular}{c r r}

\toprule

\addlinespace[-3mm]
Label & Integration over $y$ & Integration over $\cos{\theta}$ \\

\addlinespace[-2mm]

\midrule

${\cal A}$  & \scalebox{1.}{$\displaystyle \int_{-y_{\mathrm{cut}}^{}}^{y_{\mathrm{cut}}^{}} dy$} & \scalebox{1.}{$\displaystyle \int^{1-\epsilon}_{-(1-\epsilon)} d \cos{\theta}$}  \\

${\cal B}$ & \scalebox{1.}{$\displaystyle \biggl( \int_{0}^{y_{\mathrm{cut}}^{}} - \int_{-y_{\mathrm{cut}}^{}}^{0} \biggr) dy$} & \scalebox{1.}{$\displaystyle \int^{1-\epsilon}_{-(1-\epsilon)} d \cos{\theta}$}  \\

${\cal C}$ & \scalebox{1.}{$\displaystyle \int_{-y_{\mathrm{cut}}^{}}^{y_{\mathrm{cut}}^{}} dy$} & \scalebox{1.}{$\displaystyle \biggl( \int^{1-\epsilon}_{0} - \int^{0}_{-(1-\epsilon)} \biggr) d \cos{\theta}$}  \\

${\cal D}$ & \scalebox{1.}{$\displaystyle \biggl( \int_{0}^{y_{\mathrm{cut}}^{}} - \int_{-y_{\mathrm{cut}}^{}}^{0} \biggr) dy$} & \scalebox{1.}{$\displaystyle \biggl( \int^{1-\epsilon}_{0} - \int^{0}_{-(1-\epsilon)} \biggr) d \cos{\theta}$}  \\

\addlinespace[1mm]

\bottomrule

\end{tabular}
\caption{\small The 4 different approaches labelled ${\cal A}$, ${\cal B}$, ${\cal C}$ and ${\cal D}$ for performing integration over the rapidity $y$ and the polar angle $\cos{\theta}$ in the differential cross section of eq.~(\ref{eq:differentialZH}). A value of $y_{\mathrm{cut}}^{}$ is $\ln{\sqrt{s/\hat{s}}}$; see eq.~(\ref{eq:allowedregion}). A smaller value can be also chosen for $y_{\mathrm{cut}}^{}$. A value of $\epsilon$ ($0 \le \epsilon < 1$) will be determined according to experimental conditions.} 
\label{table:integration}
\end{table}
\renewcommand{\arraystretch}{1}

\renewcommand{\arraystretch}{1.5}

\begin{table}[t]
\centering
\begin{tabular}{c c c}

\midrule
$c_1^{}$ & $\frac{4}{3} -2\epsilon + \epsilon^2_{} - \frac{1}{3}\epsilon^3_{}$ & $\frac{4}{3}$  \\

$c_2^{}$ & $\frac{4}{3} -2\epsilon^2_{} + \frac{2}{3}\epsilon^3_{}$ & $\frac{4}{3}$    \\

$c_3^{}$ & $\frac{1}{\sqrt{2}} \bigl[(1-\epsilon)\sqrt{\epsilon(2-\epsilon)} + \sin^{-1}_{}{(1-\epsilon)} \bigr]$ & $\frac{\pi}{2\sqrt{2}}$   \\

$c_4^{}$ & $\frac{2}{3\sqrt{2}} \bigl[1-2 \epsilon \sqrt{\epsilon(2-\epsilon)} + \sqrt{\epsilon^5_{}(2-\epsilon)} \bigr]$ & $\frac{2}{3\sqrt{2}}$ \\

$c_5^{}$ & $1 -2\epsilon + \epsilon^2_{}$ & $1$  \\

\midrule
\end{tabular}
\caption{\small Constant coefficients in the density matrices after $\cos{\theta}$ integration; see eqs.~(\ref{eq:cosInt1}) and (\ref{eq:cosInt2}). The last column gives the values for $\epsilon =0$.} 
\label{table:coefficients}
\end{table}

\renewcommand{\arraystretch}{1}

Now we derive the differential cross sections with respect to $\hat{s}$, $\cos{\widehat{\theta}}$ and $\widehat{\phi}$ by integrating over $y$ and $\cos{\theta}$ in eq.~(\ref{eq:differentialZH}). We consider the 4 different integration approaches summarised in Table~\ref{table:integration}. 
In the complete differential cross section of eq.~(\ref{eq:differentialZH}), the polar angle $\theta$ dependence appears only in $\rho_{\sigma}^{}(q\bar{q})$ and $\rho_{\sigma}^{}(\bar{q}q)$. The $\cos{\theta}$ integration in the approaches ${\cal A}$ and ${\cal B}$ results in
\begin{subequations}\label{eq:cosInt1}
\begin{align}
\bigl\langle \rho_{\sigma}^{}(q\bar{q}) \bigr\rangle
\equiv \int^{1-\epsilon}_{-(1-\epsilon)} d \cos{\theta}\ \rho_{\sigma}^{}(q\bar{q})
=
\left(
\begin{array}{ccc}
c_1^{} \bigl| \hat{M}_{\sigma}^+ \bigr|^2_{} & \frac{1}{2} c_2^{} \bigl( \hat{M}_{\sigma}^{+} \bigr)^*_{} \hat{M}_{\sigma}^{-}  & c_3^{} \sigma \bigl( \hat{M}_{\sigma}^{+} \bigr)^*_{} \hat{M}_{\sigma}^{0} \\
\frac{1}{2} c_2^{} \hat{M}_{\sigma}^{+} \bigl( \hat{M}_{\sigma}^{-} \bigr)^*_{}  & c_1^{} \bigl| \hat{M}_{\sigma}^- \bigr|^2_{} & c_3^{} \sigma \bigl( \hat{M}_{\sigma}^{-} \bigr)^*_{} \hat{M}_{\sigma}^{0}  \\
c_3^{} \sigma \hat{M}_{\sigma}^{+} \bigl( \hat{M}_{\sigma}^{0} \bigr)^*_{} & c_3^{} \sigma \hat{M}_{\sigma}^{-} \bigl( \hat{M}_{\sigma}^{0} \bigr)^*_{} & c_2^{} \bigl| \hat{M}_{\sigma}^0 \bigr|^2_{} \\
\end{array}
\right)\label{eq:densqqbarint1}
\end{align}
and
\begin{align}
\bigl\langle \rho_{\sigma}^{}(\bar{q}q) \bigr\rangle
\equiv \int^{1-\epsilon}_{-(1-\epsilon)} d \cos{\theta}\ \rho_{\sigma}^{}(\bar{q}q)
=
\left(
\begin{array}{ccc}
c_1^{} \bigl| \hat{M}_{\sigma}^+ \bigr|^2_{} & \frac{1}{2} c_2^{} \bigl( \hat{M}_{\sigma}^{+} \bigr)^*_{} \hat{M}_{\sigma}^{-}  & - c_3^{} \sigma \bigl( \hat{M}_{\sigma}^{+} \bigr)^*_{} \hat{M}_{\sigma}^{0} \\
\frac{1}{2} c_2^{} \hat{M}_{\sigma}^{+} \bigl( \hat{M}_{\sigma}^{-} \bigr)^*_{}  & c_1^{} \bigl| \hat{M}_{\sigma}^- \bigr|^2_{} & - c_3^{} \sigma \bigl( \hat{M}_{\sigma}^{-} \bigr)^*_{} \hat{M}_{\sigma}^{0}  \\
- c_3^{} \sigma \hat{M}_{\sigma}^{+} \bigl( \hat{M}_{\sigma}^{0} \bigr)^*_{} & - c_3^{} \sigma \hat{M}_{\sigma}^{-} \bigl( \hat{M}_{\sigma}^{0} \bigr)^*_{} & c_2^{} \bigl| \hat{M}_{\sigma}^0 \bigr|^2_{} \\
\end{array}
\right), \label{eq:densqbarqint1}
\end{align}
\end{subequations}
where $c_i^{}$ $(i=1,2,3)$ are constant values depending on $\epsilon$ and summarised in Table~\ref{table:coefficients}, and $\hat{M}_{\sigma}^{\lambda=\pm,0}$ are defined in eq.~(\ref{eq:Zsubamplitudes}). 
Notice the difference between eq.~(\ref{eq:densqqbarint1}) and eq.~(\ref{eq:densqbarqint1}) that there is a minus sign in front of the elements that have the overall $\sigma$ in eq.~(\ref{eq:densqbarqint1}). This result is actually obvious from the comparison between the amplitudes in eq.~(\ref{eq:ZampAll1}) and those in eq.~(\ref{eq:ZampAll2}). 
After integration over $y$ and $\cos{\theta}$ in the approach ${\cal A}$, the differential cross section can be expressed as
\begin{align}
\frac{d\sigma}{d\hat{s}\ d\cos{\widehat{\theta}}\ d\widehat{\phi} }\biggr|_{{\cal A}}^{}
 & = {\cal T} \int_{-y_{\mathrm{cut}}^{}}^{y_{\mathrm{cut}}^{}} dy \biggl[ q(x_1^{}) \bar{q}(x_2^{})\ d_{}^{\tau \dagger} \sum_{\sigma}^{} \bigl \langle \rho_{\sigma}^{}(q\bar{q}) \bigr \rangle d_{}^{\tau}
+
\bar{q}(x_1^{}) q(x_2^{})\ d_{}^{\tau \dagger} \sum_{\sigma}^{} \bigl \langle \rho_{\sigma}^{}(\bar{q}q) \bigr \rangle d_{}^{\tau} \biggr] \nonumber \\
 & = {\cal T} \underbrace{\int_{0}^{y_{\mathrm{cut}}^{}}}_{x_1^{}>x_2^{}} dy \biggl[ \underbrace{q(x_1^{}) \bar{q}(x_2^{})}_{A}\ d_{}^{\tau \dagger} \sum_{\sigma}^{} \bigl \langle \rho_{\sigma}^{}(q\bar{q}) \bigr \rangle d_{}^{\tau}
+
\underbrace{\bar{q}(x_1^{}) q(x_2^{})}_{B}\ d_{}^{\tau \dagger} \sum_{\sigma}^{} \bigl \langle \rho_{\sigma}^{}(\bar{q}q) \bigr \rangle d_{}^{\tau} \biggr] \nonumber \\
& +
{\cal T} \underbrace{\int^{0}_{-y_{\mathrm{cut}}^{}}}_{x_2^{}>x_1^{}} dy \biggl[ \underbrace{\bar{q}(x_2^{}) q(x_1^{})}_B \ d_{}^{\tau \dagger} \sum_{\sigma}^{} \bigl \langle \rho_{\sigma}^{}(q\bar{q}) \bigr \rangle d_{}^{\tau}
+
 \underbrace{q(x_2^{}) \bar{q}(x_1^{})}_{A}\ d_{}^{\tau \dagger} \sum_{\sigma}^{} \bigl \langle \rho_{\sigma}^{}(\bar{q}q) \bigr \rangle d_{}^{\tau} \biggr] 
\nonumber \\
 =
{\cal T}  \int_{0}^{y_{\mathrm{cut}}^{}} & dy\ 2 \Bigl[  q(x_1^{}) \bar{q}(x_2^{}) +  \bar{q}(x_1^{}) q(x_2^{}) \Bigr]
d_{}^{\tau \dagger} \sum_{\sigma}^{}
\left(
\begin{array}{ccc}
c_1^{} \bigl| \hat{M}_{\sigma}^+ \bigr|^2_{} & \frac{1}{2} c_2^{} \bigl( \hat{M}_{\sigma}^{+} \bigr)^*_{} \hat{M}_{\sigma}^{-}  & 0 \\
\frac{1}{2} c_2^{} \hat{M}_{\sigma}^{+} \bigl( \hat{M}_{\sigma}^{-} \bigr)^*_{}  & c_1^{} \bigl| \hat{M}_{\sigma}^- \bigr|^2_{} & 0  \\
0 & 0 & c_2^{} \bigl| \hat{M}_{\sigma}^0 \bigr|^2_{} \\
\end{array}
\right)
d_{}^{\tau}
,\label{eq:PartdifferentialZH1}
\end{align}
where
\begin{align}
{\cal T} =\frac{m_Z^{}k}{ 12288 \pi^3_{} \Gamma_Z^{} s \hat{s}^{\frac{3}{2}}_{}} 
\sum_f \sum_{\tau} \bigl| g_{Zf\bar{f}}^{\tau} \bigr|^2_{} C_f^{} \sum_q \label{eq:simplywriting}
\end{align}
is introduced to simplify our writing. The PDFs labelled $A$ in the second equality (i.e. $q(x_1^{}) \bar{q}(x_2^{})$ and $q(x_2^{}) \bar{q}(x_1^{})$) give the same numerical contribution after integration over $y$, therefore they are combined in the third (i.e. last) equality. The same is done for the PDFs labelled $B$.
In the last equality, $\bigl \langle \rho_{\sigma}^{}(q\bar{q}) \bigr \rangle$ and $\bigl \langle \rho_{\sigma}^{}(\bar{q}q) \bigr \rangle$ are added, and the elements that have the overall $\sigma$ vanish due to the sign difference in $\bigl \langle \rho_{\sigma}^{}(q\bar{q}) \bigr \rangle$ and $\bigl \langle \rho_{\sigma}^{}(\bar{q}q) \bigr \rangle$. 
The vanished elements in the approach ${\cal A}$ in eq.~(\ref{eq:PartdifferentialZH1}) revive by performing integration in the approach ${\cal B}$:
\begin{align}
\frac{d\sigma}{d\hat{s}\ d\cos{\widehat{\theta}}\ d\widehat{\phi} }\biggr|_{{\cal B}}^{}
& =  {\cal T} \biggl( \int_{0}^{y_{\mathrm{cut}}^{}} - \int^{0}_{-y_{\mathrm{cut}}^{}} \biggr)dy \biggl[ q(x_1^{}) \bar{q}(x_2^{})\ d_{}^{\tau \dagger} \sum_{\sigma}^{} \bigl \langle \rho_{\sigma}^{}(q\bar{q}) \bigr \rangle d_{}^{\tau}
+
\bar{q}(x_1^{}) q(x_2^{})\ d_{}^{\tau \dagger} \sum_{\sigma}^{} \bigl \langle \rho_{\sigma}^{}(\bar{q}q) \bigr \rangle d_{}^{\tau} \biggr] \nonumber \\
 & = {\cal T} \underbrace{\int_{0}^{y_{\mathrm{cut}}^{}}}_{x_1^{}>x_2^{}} dy \biggl[ \underbrace{q(x_1^{}) \bar{q}(x_2^{})}_{A}\ d_{}^{\tau \dagger} \sum_{\sigma}^{} \bigl \langle \rho_{\sigma}^{}(q\bar{q}) \bigr \rangle d_{}^{\tau}
+
\underbrace{\bar{q}(x_1^{}) q(x_2^{})}_{B}\ d_{}^{\tau \dagger} \sum_{\sigma}^{} \bigl \langle \rho_{\sigma}^{}(\bar{q}q) \bigr \rangle d_{}^{\tau} \biggr] \nonumber \\
& -
{\cal T} \underbrace{\int^{0}_{-y_{\mathrm{cut}}^{}}}_{x_2^{}>x_1^{}} dy \biggl[ \underbrace{\bar{q}(x_2^{}) q(x_1^{})}_B \ d_{}^{\tau \dagger} \sum_{\sigma}^{} \bigl \langle \rho_{\sigma}^{}(q\bar{q}) \bigr \rangle d_{}^{\tau}
+
 \underbrace{q(x_2^{}) \bar{q}(x_1^{})}_{A}\ d_{}^{\tau \dagger} \sum_{\sigma}^{} \bigl \langle \rho_{\sigma}^{}(\bar{q}q) \bigr \rangle d_{}^{\tau} \biggr] 
\nonumber \\
 = 
{\cal T} \int_{0}^{y_{\mathrm{cut}}^{}} & dy\ 2 \Bigl[ q(x_1^{}) \bar{q}(x_2^{}) - \bar{q}(x_1^{}) q(x_2^{}) \Bigr]
d_{}^{\tau \dagger} \sum_{\sigma}^{}
\left(
\begin{array}{ccc}
0 & 0  & c_3^{} \sigma \bigl( \hat{M}_{\sigma}^{+} \bigr)^*_{} \hat{M}_{\sigma}^{0} \\
0  & 0 & c_3^{} \sigma \bigl( \hat{M}_{\sigma}^{-} \bigr)^*_{} \hat{M}_{\sigma}^{0}  \\
c_3^{} \sigma \hat{M}_{\sigma}^{+} \bigl( \hat{M}_{\sigma}^{0} \bigr)^*_{} & c_3^{} \sigma \hat{M}_{\sigma}^{-} \bigl( \hat{M}_{\sigma}^{0} \bigr)^*_{} & 0 \\
\end{array}
\right)
d_{}^{\tau},\label{eq:PartdifferentialZH2}
\end{align}
where, in the last equality, $\bigl \langle \rho_{\sigma}^{}(q\bar{q}) \bigr \rangle$ and $\bigl \langle \rho_{\sigma}^{}(\bar{q}q) \bigr \rangle$ are subtracted. As a result, in contrast to the approach ${\cal A}$, only the elements that have the overall $\sigma$ survive. \\

The $\cos{\theta}$ integration in the approaches ${\cal C}$ and ${\cal D}$ results in
\begin{subequations}\label{eq:cosInt2}
\begin{align}
\overline{ \bigl \langle \rho_{\sigma}^{}(q\bar{q}) \bigr \rangle } \equiv
\biggl( \int^{1-\epsilon}_{0} - \int^{0}_{-(1-\epsilon)} \biggr) d \cos{\theta}\ \rho_{\sigma}^{}(q\bar{q})
=
\left(
\begin{array}{ccc}
c_5^{} \sigma \bigl| \hat{M}_{\sigma}^+ \bigr|^2_{} & 0  & c_4^{} \bigl( \hat{M}_{\sigma}^{+} \bigr)^*_{} \hat{M}_{\sigma}^{0} \\
0  & - c_5^{} \sigma \bigl| \hat{M}_{\sigma}^- \bigr|^2_{} & - c_4^{} \bigl( \hat{M}_{\sigma}^{-} \bigr)^*_{} \hat{M}_{\sigma}^{0}  \\
c_4^{} \hat{M}_{\sigma}^{+} \bigl( \hat{M}_{\sigma}^{0} \bigr)^*_{} & - c_4^{} \hat{M}_{\sigma}^{-} \bigl( \hat{M}_{\sigma}^{0} \bigr)^*_{} & 0 \\
\end{array}
\right) \label{eq:densqqbarint2}
\end{align}
and
\begin{align}
\overline{ \bigl \langle \rho_{\sigma}^{}(\bar{q}q) \bigr \rangle } \equiv
\biggl( \int^{1-\epsilon}_{0} - \int^{0}_{-(1-\epsilon)} \biggr) d \cos{\theta}\ \rho_{\sigma}^{}(\bar{q}q)
=
\left(
\begin{array}{ccc}
- c_5^{} \sigma \bigl| \hat{M}_{\sigma}^+ \bigr|^2_{} & 0  & c_4^{} \bigl( \hat{M}_{\sigma}^{+} \bigr)^*_{} \hat{M}_{\sigma}^{0} \\
0  &  c_5^{} \sigma \bigl| \hat{M}_{\sigma}^- \bigr|^2_{} & - c_4^{} \bigl( \hat{M}_{\sigma}^{-} \bigr)^*_{} \hat{M}_{\sigma}^{0}  \\
c_4^{} \hat{M}_{\sigma}^{+} \bigl( \hat{M}_{\sigma}^{0} \bigr)^*_{} & - c_4^{} \hat{M}_{\sigma}^{-} \bigl( \hat{M}_{\sigma}^{0} \bigr)^*_{} & 0 \\
\end{array}
\right), \label{eq:densqbarqint2}
\end{align}
\end{subequations}
where the constant values $c_i^{}$ $(i=4,5)$ are summarised in Table~\ref{table:coefficients}. Notice again that the sign in front of the elements that have the overall $\sigma$ is different between $\overline{ \langle \rho_{\sigma}^{}(q\bar{q}) \rangle }$ and $\overline{ \langle \rho_{\sigma}^{}(\bar{q}q) \rangle }$. The integration over $y$ and $\cos{\theta}$ in the approaches ${\cal C}$ and ${\cal D}$ proceeds in the same manners as eq.~(\ref{eq:PartdifferentialZH1}) and eq.~(\ref{eq:PartdifferentialZH2}), respectively:
\begin{align}
\frac{d\sigma}{d\hat{s}\ d\cos{\widehat{\theta}}\ d\widehat{\phi} }\biggr|_{{\cal C}}^{}
 & = {\cal T} \int_{-y_{\mathrm{cut}}^{}}^{y_{\mathrm{cut}}^{}} dy \biggl[ q(x_1^{}) \bar{q}(x_2^{})\ d_{}^{\tau \dagger} \sum_{\sigma}^{} \overline{ \bigl \langle \rho_{\sigma}^{}(q\bar{q}) \bigr \rangle }  d_{}^{\tau}
+
\bar{q}(x_1^{}) q(x_2^{})\ d_{}^{\tau \dagger} \sum_{\sigma}^{} \overline{ \bigl \langle \rho_{\sigma}^{}(\bar{q}q) \bigr \rangle }  d_{}^{\tau} \biggr] \nonumber \\
  =
{\cal T} \int_{0}^{y_{\mathrm{cut}}^{}} & dy\ 2 \Bigl[  q(x_1^{}) \bar{q}(x_2^{}) +  \bar{q}(x_1^{}) q(x_2^{}) \Bigr]
d_{}^{\tau \dagger} \sum_{\sigma}^{}
\left(
\begin{array}{ccc}
0 & 0  & c_4^{} \bigl( \hat{M}_{\sigma}^{+} \bigr)^*_{} \hat{M}_{\sigma}^{0}  \\
0 & 0 & - c_4^{} \bigl( \hat{M}_{\sigma}^{-} \bigr)^*_{} \hat{M}_{\sigma}^{0}   \\
c_4^{} \hat{M}_{\sigma}^{+} \bigl( \hat{M}_{\sigma}^{0} \bigr)^*_{}  & - c_4^{} \hat{M}_{\sigma}^{-} \bigl( \hat{M}_{\sigma}^{0} \bigr)^*_{}  & 0 \\
\end{array}
\right)
d_{}^{\tau},
\label{eq:PartdifferentialZH3}
\end{align}
\begin{align}
\frac{d\sigma}{d\hat{s}\ d\cos{\widehat{\theta}}\ d\widehat{\phi} }\biggr|_{{\cal D}}^{}
& =  {\cal T} \biggl( \int_{0}^{y_{\mathrm{cut}}^{}} - \int^{0}_{-y_{\mathrm{cut}}^{}} \biggr)dy \biggl[ q(x_1^{}) \bar{q}(x_2^{})\ d_{}^{\tau \dagger} \sum_{\sigma}^{}  \overline{ \bigl \langle \rho_{\sigma}^{}(q\bar{q}) \bigr \rangle } d_{}^{\tau}
+
\bar{q}(x_1^{}) q(x_2^{})\ d_{}^{\tau \dagger} \sum_{\sigma}^{} \overline{ \bigl \langle \rho_{\sigma}^{}(\bar{q}q) \bigr \rangle }  d_{}^{\tau} \biggr] \nonumber \\
 = 
{\cal T} \int_{0}^{y_{\mathrm{cut}}^{}} & dy\ 2 \Bigl[ q(x_1^{}) \bar{q}(x_2^{}) - \bar{q}(x_1^{}) q(x_2^{}) \Bigr]
d_{}^{\tau \dagger} \sum_{\sigma}^{}
\left(
\begin{array}{ccc}
c_5^{} \sigma \bigl| \hat{M}_{\sigma}^+ \bigr|^2_{} & 0  & 0 \\
0  & - c_5^{} \sigma \bigl| \hat{M}_{\sigma}^- \bigr|^2_{} & 0  \\
0 & 0 & 0 \\
\end{array}
\right)
d_{}^{\tau}.\label{eq:PartdifferentialZH4}
\end{align}
Only the elements that do not have the overall $\sigma$ survive in eq.~(\ref{eq:PartdifferentialZH3}), while only the elements that have the overall $\sigma$ survive in eq.~(\ref{eq:PartdifferentialZH4}).\\

The differential cross sections with respect to $\hat{s}$ and the decay angles $\widehat{\theta}$ and $\widehat{\phi}$ have 9 independent angular distributions and can be expressed as
\begin{align}
\frac{d\sigma}{d\hat{s}\ d\cos{\widehat{\theta}}\ d\widehat{\phi} }\biggr|_{i(={\cal A, B, C, D})}^{}
 &=  F_{i1}^{} \bigl(1+\cos^2_{}{\widehat{\theta}} \bigr)
+ F_{i2}^{} \bigl(1-3\cos^2_{}{\widehat{\theta}} \bigr)
+ F_{i3}^{} \cos{\widehat{\theta}} \nonumber  \\
& + F_{i4}^{} \sin{\widehat{\theta}} \cos{\widehat{\phi}}  
+ F_{i5}^{} \sin{2\widehat{\theta}} \cos{\widehat{\phi}} 
+ F_{i6}^{} \sin^2_{}{\widehat{\theta}} \cos{2\widehat{\phi}} \nonumber \\
& + F_{i7}^{} \sin{\widehat{\theta}} \sin{\widehat{\phi}} 
+ F_{i8}^{} \sin{2\widehat{\theta}} \sin{\widehat{\phi}} 
+ F_{i9}^{} \sin^2_{}{\widehat{\theta}} \sin{2\widehat{\phi}},\label{eq:diffcrossgeneral}
\end{align}
where the coefficients $F_{ia}^{}$ $(i={\cal A, B, C, D})$ $(a=1,2,\cdots,9)$ are functions of $\hat{s}$ and written in terms of the non-vanishing elements in eqs.~(\ref{eq:PartdifferentialZH1}), (\ref{eq:PartdifferentialZH2}), (\ref{eq:PartdifferentialZH3}) and (\ref{eq:PartdifferentialZH4}). Note that there are in total 36 ($=4 \times 9$) coefficients. 
It is straightforward to obtain the explicit form of the coefficients $F_{ia}^{}$ $(i={\cal A, B, C, D})$ $(a=1,2,\cdots,9)$:
\begin{subequations}\label{eq:observables}
\begin{align}
F_{ {\cal A} ({\cal C}) a}^{}
& =
{\cal T} \int_{0}^{y_{\mathrm{cut}}^{}} dy\ 2 \Bigl[  q(x_1^{}) \bar{q}(x_2^{}) +  \bar{q}(x_1^{}) q(x_2^{}) \Bigr] \sum_{\sigma} f_{ {\cal A} ({\cal C}) a}^{}, \\
F_{ {\cal B} ({\cal D}) a}^{}
& =
{\cal T} \int_{0}^{y_{\mathrm{cut}}^{}} dy\ 2 \Bigl[  q(x_1^{}) \bar{q}(x_2^{}) -  \bar{q}(x_1^{}) q(x_2^{}) \Bigr] \sum_{\sigma} f_{ {\cal B} ({\cal D}) a}^{},
\end{align}
\end{subequations}
where
\begin{subequations}\label{eq:observables-small}
\begin{align}
f_{{\cal A}1}^{} & = \frac{1}{2}\bigl( c_1^{} \bigl| \hat{M}_{\sigma}^+ \bigr|^2_{} + c_1^{} \bigl| \hat{M}_{\sigma}^- \bigr|^2_{} + c_2^{} \bigl| \hat{M}_{\sigma}^0 \bigr|^2_{} \bigr),
& f_{{\cal B}1}^{} & = 0, \nonumber \\
f_{{\cal A}2}^{} & = \frac{1}{2} c_2^{} \bigl| \hat{M}_{\sigma}^0 \bigr|^2_{},
& f_{{\cal B}2}^{} & = 0, \nonumber \\
f_{{\cal A}3}^{} & = c_1^{} \bigl( \bigl| \hat{M}_{\sigma}^+ \bigr|^2_{} - \bigl| \hat{M}_{\sigma}^- \bigr|^2_{} \bigr)\tau,
& f_{{\cal B}3}^{} & = 0, \nonumber \\
f_{{\cal A}4}^{} & = 0,
& f_{{\cal B}4}^{} & = \sqrt{2} \sigma c_3^{} {\it Re} \bigl[ \bigl( \hat{M}_{\sigma}^{+} \bigr)^*_{} \hat{M}_{\sigma}^{0} + \bigl( \hat{M}_{\sigma}^{0} \bigr)^*_{} \hat{M}_{\sigma}^{-} \bigr] \tau, \nonumber \\
f_{{\cal A}5}^{} & = 0,
& f_{{\cal B}5}^{} & = \frac{1}{\sqrt{2}} \sigma c_3^{} {\it Re} \bigl[ \bigl( \hat{M}_{\sigma}^{+} \bigr)^*_{} \hat{M}_{\sigma}^{0} - \bigl( \hat{M}_{\sigma}^{0} \bigr)^*_{} \hat{M}_{\sigma}^{-} \bigr] , \nonumber \\
f_{{\cal A}6}^{} & = \frac{1}{2} c_2^{} {\it Re} \bigl[ \bigl( \hat{M}_{\sigma}^{+} \bigr)^*_{} \hat{M}_{\sigma}^{-} \bigr] ,
& f_{{\cal B}6}^{} & = 0, \nonumber \\
f_{{\cal A}7}^{} & = 0,
& f_{{\cal B}7}^{} & = \sqrt{2} \sigma c_3^{} {\it Im} \bigl[ \bigl( \hat{M}_{\sigma}^{+} \bigr)^*_{} \hat{M}_{\sigma}^{0} + \bigl( \hat{M}_{\sigma}^{0} \bigr)^*_{} \hat{M}_{\sigma}^{-} \bigr] \tau, \nonumber \\
f_{{\cal A}8}^{} & = 0,
& f_{{\cal B}8}^{} & = \frac{1}{\sqrt{2}} \sigma c_3^{} {\it Im} \bigl[ \bigl( \hat{M}_{\sigma}^{+} \bigr)^*_{} \hat{M}_{\sigma}^{0} - \bigl( \hat{M}_{\sigma}^{0} \bigr)^*_{} \hat{M}_{\sigma}^{-} \bigr] , \nonumber \\
f_{{\cal A}9}^{} & = \frac{1}{2} c_2^{} {\it Im} \bigl[ \bigl( \hat{M}_{\sigma}^{+} \bigr)^*_{} \hat{M}_{\sigma}^{-} \bigr] ,
& f_{{\cal B}9}^{} & = 0,
\end{align}
and
\begin{align}
f_{{\cal C}1}^{} & = 0,
& f_{{\cal D}1}^{} & = \frac{1}{2} \sigma c_5^{} \bigl( \bigl| \hat{M}_{\sigma}^+ \bigr|^2_{} - \bigl| \hat{M}_{\sigma}^- \bigr|^2_{} \bigr), \nonumber \\
f_{{\cal C}2}^{} & = 0,
& f_{{\cal D}2}^{} & = 0, \nonumber \\
f_{{\cal C}3}^{} & = 0,
& f_{{\cal D}3}^{} & = \sigma c_5^{} \bigl( \bigl| \hat{M}_{\sigma}^+ \bigr|^2_{} + \bigl| \hat{M}_{\sigma}^- \bigr|^2_{} \bigr) \tau, \nonumber \\
f_{{\cal C}4}^{} & = \sqrt{2}  c_4^{} {\it Re} \bigl[ \bigl( \hat{M}_{\sigma}^{+} \bigr)^*_{} \hat{M}_{\sigma}^{0} - \bigl( \hat{M}_{\sigma}^{0} \bigr)^*_{} \hat{M}_{\sigma}^{-} \bigr] \tau,
& f_{{\cal D}4}^{} & = 0, \nonumber \\
f_{{\cal C}5}^{} & = \frac{1}{\sqrt{2}} c_4^{} {\it Re} \bigl[ \bigl( \hat{M}_{\sigma}^{+} \bigr)^*_{} \hat{M}_{\sigma}^{0} + \bigl( \hat{M}_{\sigma}^{0} \bigr)^*_{} \hat{M}_{\sigma}^{-} \bigr],
& f_{{\cal D}5}^{} & = 0, \nonumber \\
f_{{\cal C}6}^{} & = 0,
& f_{{\cal D}6}^{} & = 0, \nonumber \\
f_{{\cal C}7}^{} & = \sqrt{2}  c_4^{} {\it Im} \bigl[ \bigl( \hat{M}_{\sigma}^{+} \bigr)^*_{} \hat{M}_{\sigma}^{0} - \bigl( \hat{M}_{\sigma}^{0} \bigr)^*_{} \hat{M}_{\sigma}^{-} \bigr] \tau,
& f_{{\cal D}7}^{} & = 0, \nonumber \\
f_{{\cal C}8}^{} & = \frac{1}{\sqrt{2}} c_4^{} {\it Im} \bigl[ \bigl( \hat{M}_{\sigma}^{+} \bigr)^*_{} \hat{M}_{\sigma}^{0} + \bigl( \hat{M}_{\sigma}^{0} \bigr)^*_{} \hat{M}_{\sigma}^{-} \bigr],
& f_{{\cal D}8}^{} & = 0, \nonumber \\
f_{{\cal C}9}^{} & = 0,
& f_{{\cal D}9}^{} & = 0.
\end{align}
\end{subequations}
Among the 36 coefficients, only the 15 coefficients can be non-zero. It is easy to notice that there are the 10 combinations of the elements of the density matrix in total. However, only 9 of them are independent~\footnote{Among the 3 combinations in $f_{{\cal A}1}^{}$, $f_{{\cal A}2}^{}$ and $f_{{\cal D}3}^{}$, only 2 of them are independent, since one of them can be constructed from the other two.}. 
Some of them are strictly zero, if the amplitudes satisfy the restriction in eq.~(\ref{eq:CPinvZ-2}) from CP invariance and/or the restriction in eq.~(\ref{eq:CPTinvZ-2}) from $\mathrm{CP\widetilde{T}}$ invariance. These symmetry properties can be explicitly checked in the following way. 
By applying the restrictions in eqs.~(\ref{eq:CPinvZ-2}) and (\ref{eq:CPTinvZ-2}) to the combination in $f_{{\cal A}9}^{}$, we find
\begin{align}
\mathrm{CP}\ \mathrm{invariance}:\ & {\it Im} \bigl[ \bigl( \hat{M}_{\sigma}^{+} \bigr)^*_{} \hat{M}_{\sigma}^{-} \bigr] 
= {\it Im} \bigl[ | \hat{M}_{\sigma}^{+} |^2_{} \bigr]  
= 0, \nonumber \\
\mathrm{CP\widetilde{T}}\ \mathrm{invariance}:\ & {\it Im} \bigl[ \bigl( \hat{M}_{\sigma}^{+} \bigr)^*_{} \hat{M}_{\sigma}^{-} \bigr] 
= {\it Im} \bigl[ ( \hat{M}_{\sigma}^{-} )^2_{} \bigr]  
\ne 0,
\end{align}
where the former means that CP invariance requires it to be zero, while the latter means that $\mathrm{CP\widetilde{T}}$ invariance does not require it to be zero. This indicates that observation of a nonzero value in $F_{ {\cal A}9}^{}$ signals CP violation. Similarly, by applying the restrictions in eqs.~(\ref{eq:CPinvZ-2}) and (\ref{eq:CPTinvZ-2}) to the combination in $f_{{\cal B}8}^{}$ and $f_{{\cal C}7}^{}$, we find
\begin{align}
\mathrm{CP}\ \mathrm{invariance}:\ & {\it Im} \bigl[ \bigl( \hat{M}_{\sigma}^{+} \bigr)^*_{} \hat{M}_{\sigma}^{0} - \bigl( \hat{M}_{\sigma}^{0} \bigr)^*_{} \hat{M}_{\sigma}^{-} \bigr]
= {\it Im} \bigl\{ 2 i {\it Im}\bigl[ \bigl( \hat{M}_{\sigma}^{+} \bigr)^*_{} \hat{M}_{\sigma}^{0} \bigr]  \bigr\}  \ne 0,\nonumber \\
\mathrm{CP\widetilde{T}}\ \mathrm{invariance}:\ & {\it Im} \bigl[ \bigl( \hat{M}_{\sigma}^{+} \bigr)^*_{} \hat{M}_{\sigma}^{0} - \bigl( \hat{M}_{\sigma}^{0} \bigr)^*_{} \hat{M}_{\sigma}^{-} \bigr]
= {\it Im} \bigl[ \bigl( \hat{M}_{\sigma}^{+} \bigr)^*_{} \hat{M}_{\sigma}^{0} - \hat{M}_{\sigma}^{0} \bigl( \hat{M}_{\sigma}^{+} \bigr)^*_{} \bigr] = 0,
\end{align}
where $\mathrm{CP\widetilde{T}}$ invariance requires it to be zero, while CP invariance does not. This indicates that observation of a nonzero value in $F_{ {\cal B}8}^{}$ or $F_{ {\cal C}7}^{}$ signals $\mathrm{CP\widetilde{T}}$ violation.
Finally, by applying the restrictions in eqs.~(\ref{eq:CPinvZ-2}) and (\ref{eq:CPTinvZ-2}) to the combination in $f_{{\cal B}5}^{}$ and $f_{{\cal C}4}^{}$, we find
\begin{align}
\mathrm{CP}\ \mathrm{invariance}:\ & {\it Re} \bigl[ \bigl( \hat{M}_{\sigma}^{+} \bigr)^*_{} \hat{M}_{\sigma}^{0} - \bigl( \hat{M}_{\sigma}^{0} \bigr)^*_{} \hat{M}_{\sigma}^{-} \bigr]
= {\it Re} \bigl\{ 2 i {\it Im}\bigl[ \bigl( \hat{M}_{\sigma}^{+} \bigr)^*_{} \hat{M}_{\sigma}^{0} \bigr]  \bigr\} = 0,\nonumber \\
\mathrm{CP\widetilde{T}}\ \mathrm{invariance}:\ & {\it Re} \bigl[ \bigl( \hat{M}_{\sigma}^{+} \bigr)^*_{} \hat{M}_{\sigma}^{0} - \bigl( \hat{M}_{\sigma}^{0} \bigr)^*_{} \hat{M}_{\sigma}^{-} \bigr]
= {\it Re} \bigl[ \bigl( \hat{M}_{\sigma}^{+} \bigr)^*_{} \hat{M}_{\sigma}^{0} - \hat{M}_{\sigma}^{0} \bigl( \hat{M}_{\sigma}^{+} \bigr)^*_{} \bigr] = 0,
\end{align}
where both CP invariance and $\mathrm{CP\widetilde{T}}$ invariance require it to be zero. This indicates that observation of a nonzero value in $F_{ {\cal B}5}^{}$ or $F_{ {\cal C}4}^{}$ signals both CP violation {\it and} $\mathrm{CP\widetilde{T}}$ violation. These symmetry properties of the 10 combinations are summarised in Table~\ref{table:symproperty}. The symbol $+$ means that the symmetry does not require the combination to be zero, while $-$ means that the symmetry requires the combination to be zero. 
We also show the coefficients $F_{ia}^{}$ by measuring which the combinations of the density matrix elements can be determined. 
The symbol $\circ$ in the column "$f$ charge" means that observation of the coefficient $F_{ia}^{}$ requires distinguishing the fermion $f$ from the antifermion $\bar{f}$. In other words, it requires identification of the charge (flavor) of the final fermion $f$. 
This can be confirmed by performing a translation $\widehat{\theta} \to \pi - \widehat{\theta}$ {\it and} $\widehat{\phi} \to \widehat{\phi} + \pi$ in eq.~(\ref{eq:diffcrossgeneral}) and observing the change of the sign. 
Among the 15 coefficients, only 9 of them do not require the charge identification of the final fermion $f$. It should be emphasised that these 9 coefficients are necessary and sufficient to determine all of the 9 independent combinations of the density matrix elements. \\

\begin{table}
\begin{tabular}{@{}l cccccc@{}} 
\toprule

\addlinespace[1mm]

Combinations of  & \multicolumn{2}{l}{Symmetry properties} & Observables & $f$ charge  & $HZ\gamma$ & d.o.p \\  

\cmidrule(l){2-3}

the density matrix elements &  CP & $\mathrm{CP\widetilde{T}}$ & & & \\ 

\midrule

\addlinespace[1mm]

$c_1^{} \bigl| \hat{M}_{\sigma}^+ \bigr|^2_{} + c_1^{} \bigl| \hat{M}_{\sigma}^- \bigr|^2_{} + c_2^{} \bigl| \hat{M}_{\sigma}^0 \bigr|^2_{}$ & $+$ & $+$  &  $F_{ {\cal A} 1}^{}$ & - & - & -\\

$\bigl| \hat{M}_{\sigma}^0 \bigr|^2_{}$ & $+$ & $+$  &  $F_{ {\cal A} 2}^{}$ & - & -  & -\\

$\bigl| \hat{M}_{\sigma}^+ \bigr|^2_{} + \bigl| \hat{M}_{\sigma}^- \bigr|^2_{}$ & $+$ & $+$  &  $F_{ {\cal D} 3}^{}$ & $\circ$ & $\circ$ & $\circ$ \\

$\bigl| \hat{M}_{\sigma}^+ \bigr|^2_{} - \bigl| \hat{M}_{\sigma}^- \bigr|^2_{}$  & $-$ & $-$  &  $F_{ {\cal A} 3}^{}$ & $\circ$ & - & -\\

 & & & $F_{ {\cal D} 1}^{}$ & - & $\circ$ & $\circ$ \\

${\it Re} \bigl[ \bigl( \hat{M}_{\sigma}^{+} \bigr)^*_{} \hat{M}_{\sigma}^{0} + \bigl( \hat{M}_{\sigma}^{0} \bigr)^*_{} \hat{M}_{\sigma}^{-} \bigr]$  & $+$ & $+$  &  $F_{ {\cal B} 4}^{}$ & $\circ$ & $\circ$ & $\circ$\\

 & & & $F_{ {\cal C} 5}^{}$ & - & - & -\\

${\it Re} \bigl[ \bigl( \hat{M}_{\sigma}^{+} \bigr)^*_{} \hat{M}_{\sigma}^{0} - \bigl( \hat{M}_{\sigma}^{0} \bigr)^*_{} \hat{M}_{\sigma}^{-} \bigr]$  & $-$ & $-$  &  $F_{ {\cal B} 5}^{}$ & - & $\circ$ & $\circ$ \\

 & & & $F_{ {\cal C} 4}^{}$ & $\circ$ & - & -\\

${\it Re} \bigl[ \bigl( \hat{M}_{\sigma}^{+} \bigr)^*_{} \hat{M}_{\sigma}^{-} \bigr]$ & $+$ & $+$  &  $F_{ {\cal A} 6}^{}$ & -  & - & - \\

${\it Im} \bigl[ \bigl( \hat{M}_{\sigma}^{+} \bigr)^*_{} \hat{M}_{\sigma}^{0} + \bigl( \hat{M}_{\sigma}^{0} \bigr)^*_{} \hat{M}_{\sigma}^{-} \bigr]$  & $-$ & $+$  &  $F_{ {\cal B} 7}^{}$ & $\circ$ & $\circ$ & $\circ$\\

 & & & $F_{ {\cal C} 8}^{}$ & - & - & - \\

${\it Im} \bigl[ \bigl( \hat{M}_{\sigma}^{+} \bigr)^*_{} \hat{M}_{\sigma}^{0} - \bigl( \hat{M}_{\sigma}^{0} \bigr)^*_{} \hat{M}_{\sigma}^{-} \bigr]$  & $+$ & $-$  &  $F_{ {\cal B} 8}^{}$ & - & $\circ$ & $\circ$\\
 & & & $F_{ {\cal C} 7}^{}$ & $\circ$ & - & - \\

${\it Im} \bigl[ \bigl( \hat{M}_{\sigma}^{+} \bigr)^*_{} \hat{M}_{\sigma}^{-} \bigr]$ & $-$ & $+$  &  $F_{ {\cal A} 9}^{}$ & - & - & -\\

\addlinespace[1mm]

\bottomrule

\end{tabular}
\caption{\small Symmetry properties of the 10 combinations of the density matrix elements. The coefficients $F_{ia}^{}$ of the differential angular distributions, by measuring which the combinations of the density matrix elements can be determined, are also shown. 
The symbol $-$ means that the symmetry (CP or $\mathrm{CP\widetilde{T}}$) requires the combination to be zero, while the symbol $+$ means that the symmetry does not; observation of a non-zero value in the combination with the symbol $-$ under CP, for instance, signals CP violation. 
The symbol $\circ$ in the column "$f$ charge" means that observation of the coefficient $F_{ia}^{}$ requires the charge (or flavor) identification of the final fermion $f$. 
The symbol $\circ$ in the column "$HZ\gamma$" indicates that the coefficient $F_{ia}^{}$ has a good sensitivity to the $HZ\gamma$ coupling; see a discussion at the paragraph of eq.~(\ref{eq:amp:simplified}). 
The symbol $\circ$ in the last column indicates that the coefficient $F_{ia}^{}$ is weakened according to the degree of polarisation of the $Z$ boson; see a discussion at the last paragraph of Section~\ref{sec:angcoeff-W}. 
By the simple replacements $\hat{M}^{\lambda}_{\sigma} \to \hat{N}^{\lambda}_{}$ and $F_{ia}^{} \to F_{ia}^{W}$, the table corresponding to the process $pp\to W^{\pm}_{}H$ is immediately obtained; see Section~\ref{sec:polWboson}.}
\label{table:symproperty}
\end{table}

The coefficients $F_{ia}^{}$ $(i={\cal A, B, C, D})$ $(a=1,2,\cdots,9)$ will be experimentally determined by measuring the decay angles $\widehat{\theta}$ and $\widehat{\phi}$ in the rest frame of the $Z$ boson (see eq.~(\ref{eq:decaykinematics})), since we completely know the differential angular distributions as eq.~(\ref{eq:diffcrossgeneral}). We just do not know the coefficients $F_{ia}^{}$ which uniquely depend on a state of polarisation of the $Z$ boson. 
With appropriate integration over $\cos{\widehat{\theta}}$ and $\widehat{\phi}$, it is possible to isolate the angular distributions: 
\begin{subequations}\label{eq:cosIntall}
\begin{align}
\int^1_{-1} d\cos{\widehat{\theta}} \frac{d\sigma}{d\hat{s} d\cos{\widehat{\theta}} d\widehat{\phi} } & =  \frac{8}{3}F_{1}^{}
+ \frac{\pi}{2} F_{4}^{} \cos{\widehat{\phi}}  
+ \frac{4}{3} F_{6}^{} \cos{2\widehat{\phi}} 
+ \frac{\pi}{2} F_{7}^{} \sin{\widehat{\phi}} 
+ \frac{4}{3} F_{9}^{} \sin{2\widehat{\phi}}, \label{eq:cosInt21}\\
\biggl( \int^1_{0} - \int_{-1}^0 \biggr) d\cos{\widehat{\theta}} \frac{d\sigma}{d\hat{s} d\cos{\widehat{\theta}} d\widehat{\phi} } & =  F_{3}^{}
+ \frac{4}{3} F_{5}^{} \cos{\widehat{\phi}}  
+ \frac{4}{3} F_{8}^{} \sin{\widehat{\phi}}, \label{eq:cosInt22}
\end{align}
\end{subequations}
and
\begin{subequations}\label{eq:phiIntall}
\begin{align}
\frac{1}{2\pi}\int^{2\pi}_{0} d\widehat{\phi} \frac{d\sigma}{d\hat{s} d\cos{\widehat{\theta}} d\widehat{\phi} } & =  F_{1}^{} \bigl(1+\cos^2_{}{\widehat{\theta}} \bigr)
+ F_{2}^{} \bigl(1-3\cos^2_{}{\widehat{\theta}} \bigr)
+ F_{3}^{} \cos{\widehat{\theta}}, \label{eq:phiInt1} \\
\frac{1}{4} \biggl( \int^{\pi/2}_{0} - \int^{\pi}_{\pi/2} - \int^{3\pi/2}_{\pi} + \int_{3\pi/2}^{2\pi} \biggr) d\widehat{\phi} \frac{d\sigma}{d\hat{s} d\cos{\widehat{\theta}} d\widehat{\phi} } & =
 F_{4}^{} \sin{\widehat{\theta}} +
 F_{5}^{} \sin{2\widehat{\theta}}, \label{eq:phiInt2} \\
\frac{1}{4} \biggl( \int^{\pi/4}_{0} - \int^{\pi/2}_{\pi/4} - \int^{3\pi/4}_{\pi/2} + \int_{3\pi/4}^{\pi} + \int_{\pi}^{5\pi/4} - \int_{5\pi/4}^{3\pi/2} & - \int_{3\pi/2}^{7\pi/4} + \int_{7\pi/4}^{2\pi} \biggr)  d\widehat{\phi} \frac{d\sigma}{d\hat{s} d\cos{\widehat{\theta}} d\widehat{\phi} }  =
 F_{6}^{} \sin^2_{}{\widehat{\theta}}, \label{eq:phiInt3}\\
\frac{1}{4} \biggl( \int^{\pi}_{0} - \int_{\pi}^{2\pi} \biggr) d\widehat{\phi} \frac{d\sigma}{d\hat{s} d\cos{\widehat{\theta}} d\widehat{\phi} } & =
 F_{7}^{} \sin{\widehat{\theta}} +
 F_{8}^{} \sin{2\widehat{\theta}}, \\
\frac{1}{4} \biggl( \int^{\pi/2}_{0} - \int^{\pi}_{\pi/2} + \int^{3\pi/2}_{\pi} - \int_{3\pi/2}^{2\pi} \biggr) d\widehat{\phi} \frac{d\sigma}{d\hat{s} d\cos{\widehat{\theta}} d\widehat{\phi} } & =
 F_{9}^{} \sin^2_{}{\widehat{\theta}}. \label{eq:phiInt5}
\end{align}
\end{subequations}
Here the index $i(={\cal A, B, C, D})$ is omitted. By combining the 2 approaches in eq.~(\ref{eq:cosIntall}) and the 5 approaches in eq.~(\ref{eq:phiIntall}), we obtain the $10$ ($=2\times 5$) combinations. The 2 of them simply give zero (i.e. eqs.~(\ref{eq:cosInt22}) and (\ref{eq:phiInt3}), and eqs.~(\ref{eq:cosInt22}) and (\ref{eq:phiInt5})). Each of the remaining 8 combinations gives one of $F_a^{}$ $(a=1,3,4,5,6,7,8,9)$. For example, eqs.~(\ref{eq:cosInt22}) and (\ref{eq:phiInt2}) gives $F_5^{}$. Only $F_2^{}$ is not determined in this method. By a fitting procedure in the differential cross section with respect to $\hat{s}$ and $\cos{\widehat{\theta}}$ in eq.~(\ref{eq:phiInt1}), $F_2^{}$ may be determined.

\begin{figure}[t]
\centering
\includegraphics[scale=0.42]{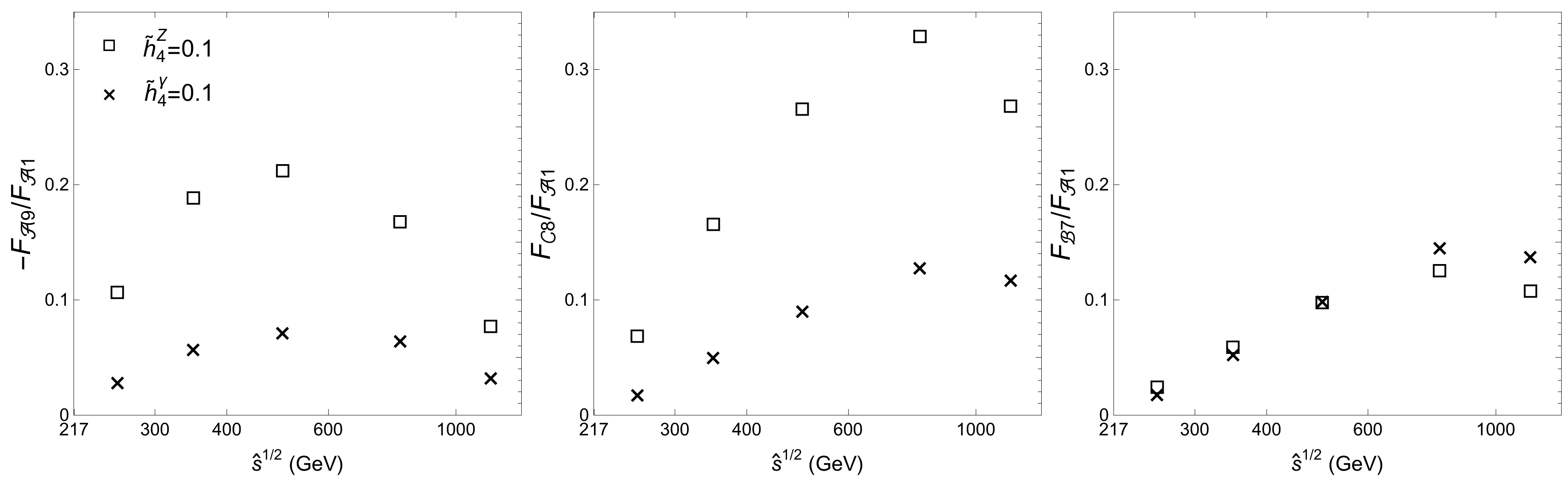}
\caption{\small 
The coefficients $-F_{{\cal A} 9}^{}$ (left panel), $F_{{\cal C} 8}^{}$ (middle panel) and $F_{{\cal B} 7}^{}$ (right panel) divided by $F_{{\cal A} 1}^{}$ are shown. These coefficients are constrained to be identically zero by CP invariance. In each panel, $\Box$ points give predictions for $\widetilde{h}_4^Z=0.1$, and $\times$ points give those for $\widetilde{h}_4^{\gamma}=0.1$. For each result, the values in the 5 bins, from left to right, are obtained after integration over $\hat{s}$ in the regions $(m_Z^{}+m_H^{}) < \hat{s}^{1/2}_{} < 300$, $300 < \hat{s}^{1/2}_{} < 400$, $400 < \hat{s}^{1/2}_{} < 600$, $600 < \hat{s}^{1/2}_{} < 1000$ and $1000 < \hat{s}^{1/2}_{} < 14000$ in units of GeV.}
\label{figure:ZHplot1}
\end{figure}

\begin{figure}[t]
\centering
\includegraphics[scale=0.42]{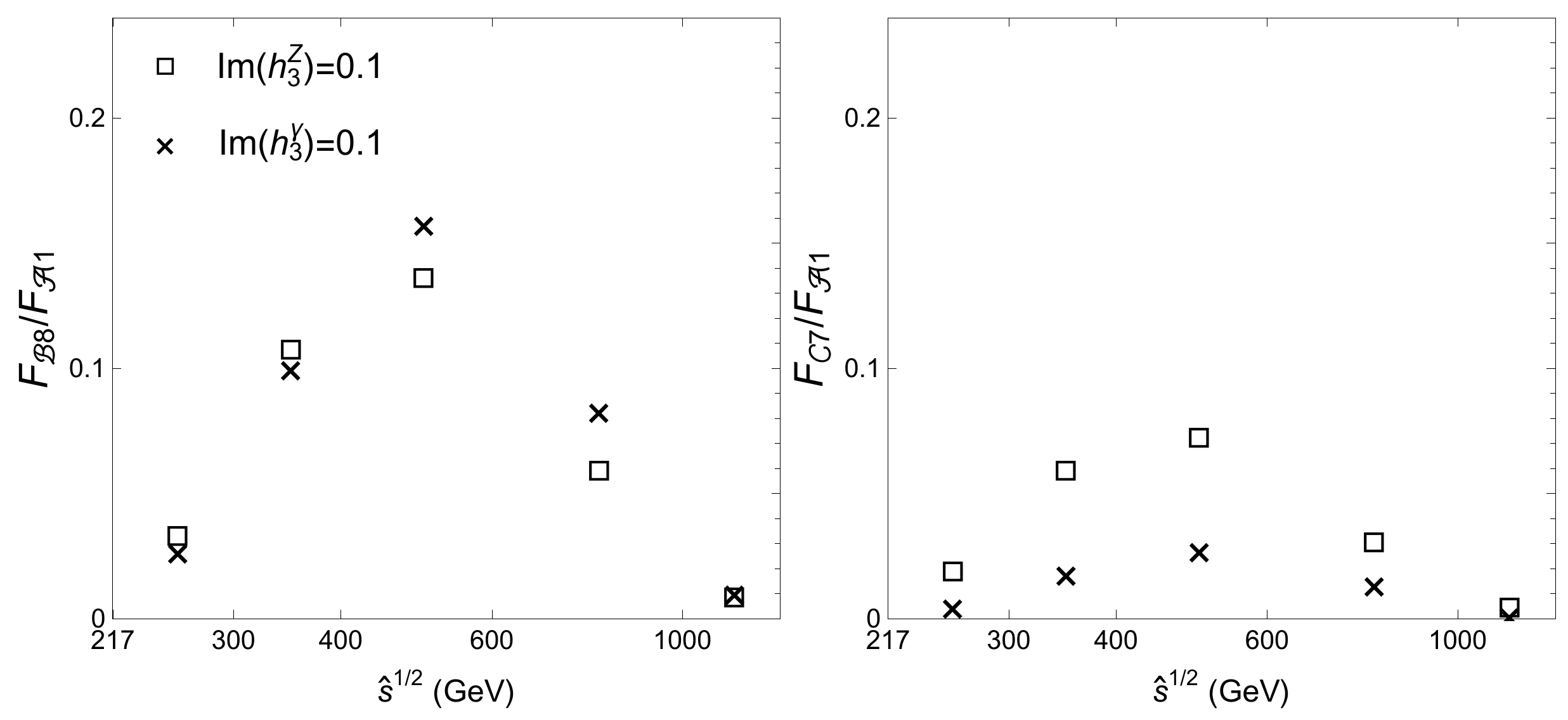}
\caption{\small 
The coefficients $F_{{\cal B} 8}^{}$ (left panel) and $F_{{\cal C} 7}^{}$ (right panel) divided by $F_{{\cal A} 1}^{}$ are shown in the same manner as Figure~\ref{figure:ZHplot1}. These coefficients are constrained to be identically zero by $\mathrm{CP\widetilde{T}}$ invariance. In each panel, $\Box$ points give predictions for $h_3^Z=0+0.1i$, and $\times$ points give those for $h_3^{\gamma}=0+0.1i$.}
\label{figure:ZHplot2}
\end{figure}

\begin{figure}[t]
\centering
\includegraphics[scale=0.35]{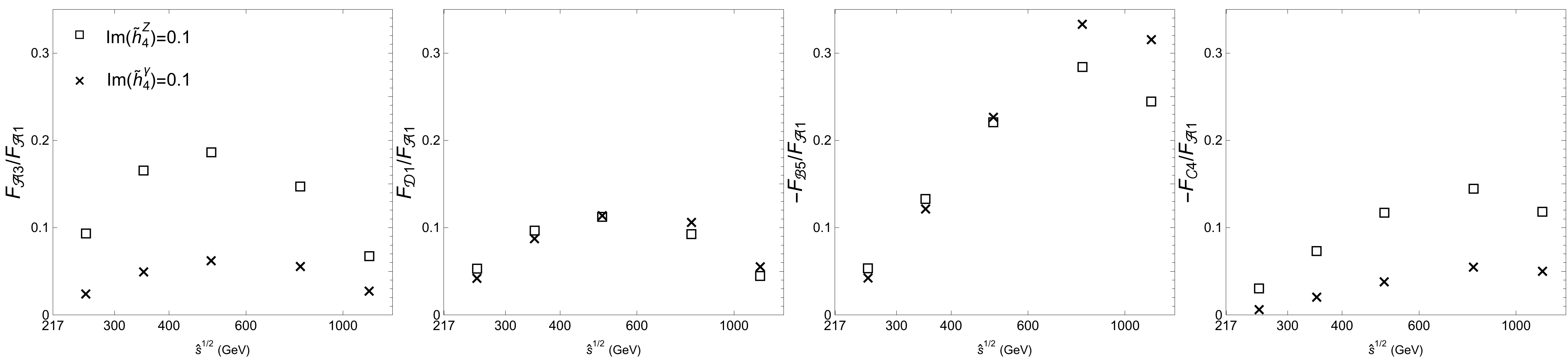}
\caption{\small 
The angular coefficients $F_{{\cal A} 3}^{}$ (left panel), $F_{{\cal D} 1}^{}$ (middle-left panel), $-F_{{\cal B} 5}^{}$ (middle-right panel) and $-F_{{\cal C} 4}^{}$ (right panel) divided by $F_{{\cal A} 1}^{}$ are shown in the same manner as Figure~\ref{figure:ZHplot1}. These coefficients are doubly constrained to be identically zero by CP invariance {\it and} $\mathrm{CP\widetilde{T}}$ invariance. In each panel, $\Box$ points give predictions for $\widetilde{h}_4^Z=0+0.1i$ and $\times$ points give those for $\widetilde{h}_4^{\gamma}=0+0.1i$.}
\label{figure:ZHplot3}
\end{figure}

\subsection{Influences of non-standard $HZZ$ and $HZ\gamma$ interactions}\label{sec:angcoeff}

In the previous section, we have clarified the restrictions on the coefficients of the differential angular distributions imposed by the CP and $\mathrm{CP\widetilde{T}}$ symmetries; see Table~\ref{table:symproperty}.  
Some of the coefficients are strictly zero in the SM due to CP invariance and some of them are small in the SM due to the smallness of re-scattering effects. These coefficients are in particular interesting as observables at the LHC, since observation of a non-zero or large value in these coefficients immediately signals the existence of physics beyond the SM. In this Section, we focus on these coefficients and study the influences of the non-standard $HZZ$ and $HZ\gamma$ couplings.\\

Our numerical results are produced for the $14$ TeV LHC. We set $m_H^{}=125.5$ GeV, and $\epsilon =0$ in the $\cos{\theta}$ integration; see eqs.~(\ref{eq:cosInt1}) and (\ref{eq:cosInt2}), and Table~\ref{table:coefficients}. 
As the final fermion flavor $f$ summed in eq.~(\ref{eq:simplywriting}), only the electron and the muon are considered when calculating the coefficients $F_{ia}^{}$ $(i={\cal A, B, C, D})$ $(a=3,4,7)$ (Recall that observation of these coefficients requires the charge identification of $f$), and all the quark flavors but the top quark are additionally considered when calculating the other coefficients. 
MSTW PDFs~\cite{Martin:2009iq} are used. The phase space integration is performed with the program BASES~\cite{Kawabata:1995th}. \\

In Figure~\ref{figure:ZHplot1}, the coefficients $-F_{{\cal A} 9}^{}$ (left panel), $F_{{\cal C} 8}^{}$ (middle panel) and $F_{{\cal B} 7}^{}$ (right panel) divided by $F_{{\cal A} 1}^{}$ are shown~\footnote{When calculating $F_{{\cal A} 1}^{}$, by which $F_{ia}^{}$ $(i={\cal A, B, C, D})$ $(a=3,4,7)$ are divided, we consider only the electron and the muon as the final fermion flavor $f$ summed in eq.~(\ref{eq:simplywriting}).}. For each result, the values in the 5 bins, from left to right, are obtained after integration over $\hat{s}$ in the regions $(m_Z^{}+m_H^{}) < \hat{s}^{1/2}_{} < 300$, $300 < \hat{s}^{1/2}_{} < 400$, $400 < \hat{s}^{1/2}_{} < 600$, $600 < \hat{s}^{1/2}_{} < 1000$ and $1000 < \hat{s}^{1/2}_{} < 14000$ in units of GeV. 
In each panel, $\Box$ points give predictions for the CP-odd form factor $\widetilde{h}_4^Z=0.1$, and $\times$ points give those for the CP-odd form factor $\widetilde{h}_4^{\gamma}=0.1$. CP invariance requires these 4 coefficients to be identically zero, thus observation of a non-zero value signals CP violation. \\

The coefficients $F_{{\cal C} 8}^{}$ and $F_{{\cal B} 7}^{}$ measure the same combination of the density matrix elements (see Table~\ref{table:symproperty}), thus show the same dependence on $\hat{s}^{1/2}_{}$. 
However, $F_{{\cal C} 8}^{}$'s sensitivity to $\widetilde{h}_4^Z$ is stronger than its sensitivity to $\widetilde{h}_4^{\gamma}$, while $F_{{\cal B} 7}^{}$'s sensitivity to $\widetilde{h}_4^Z$ is comparable to its sensitivity to $\widetilde{h}_4^{\gamma}$. 
This difference originates from the overall $\sigma$ in $f_{{\cal B} 7}^{}$ and is explained as follows. Eq.~(\ref{eq:Zsubamplitudes}) shows that $\hat{M}^{\lambda}_+$ and $\hat{M}^{\lambda}_-$ can be written in the following simplified form as the sum of the $Z$ boson contribution and the $\gamma$ contribution:
\begin{subequations}\label{eq:amp:simplified}
\begin{align}
\hat{M}^{\lambda}_+ = &  g_{Z+}^{} f_Z^{\lambda} + f_{\gamma}^{\lambda},\\
\hat{M}^{\lambda}_- = &  g_{Z-}^{} f_Z^{\lambda} + f_{\gamma}^{\lambda}.
\end{align}
\end{subequations}
By using this expression, we derive
\begin{subequations}
\begin{align}
(\hat{M}^{\lambda^{\prime}_{}}_+)^*_{} \times \hat{M}^{\lambda}_+
=
g_{Z+}^2 (f_Z^{\lambda^{\prime}_{}})^*_{} f_Z^{\lambda}
+
g_{Z+}^{} \bigl[ (f_Z^{\lambda^{\prime}_{}})^*_{} f_{\gamma}^{\lambda} + (f_{\gamma}^{\lambda^{\prime}_{}})^*_{} f_{Z}^{\lambda} \bigr]
+
(f_{\gamma}^{\lambda^{\prime}_{}})^*_{} f_{\gamma}^{\lambda}, \label{eq:interfer1} \\
(\hat{M}^{\lambda^{\prime}_{}}_-)^*_{} \times \hat{M}^{\lambda}_-
=
g_{Z-}^2 (f_Z^{\lambda^{\prime}_{}})^*_{} f_Z^{\lambda}
+
g_{Z-}^{} \bigl[ (f_Z^{\lambda^{\prime}_{}})^*_{} f_{\gamma}^{\lambda} + (f_{\gamma}^{\lambda^{\prime}_{}})^*_{} f_{Z}^{\lambda} \bigr]
+
(f_{\gamma}^{\lambda^{\prime}_{}})^*_{} f_{\gamma}^{\lambda}. \label{eq:interfer2}
\end{align}
\end{subequations}
The couplings $g_{Z+}^{}$ and $g_{Z-}^{}$ have opposite signs to each other, both for the up-type quarks and for the down-type quarks. 
The interference term between the $Z$ boson contribution and the $\gamma$ contribution in eq.~(\ref{eq:interfer1}) and that in eq.~(\ref{eq:interfer2}), therefore, have opposite signs to each other, too. Considering $f_{\gamma}^{\lambda}=0$ at the tree level approximation in the SM, the interference terms may give a larger contribution than the $(f_{\gamma}^{\lambda^{\prime}_{}})^*_{} f_{\gamma}^{\lambda}$ term. 
The coefficient $F_{{\cal C} 8}^{}$ has a structure that eq.~(\ref{eq:interfer1}) and eq.~(\ref{eq:interfer2}) are added and the interference terms tend to cancel to each other. On the other hand, $F_{{\cal B} 7}^{}$ has a structure that eq.~(\ref{eq:interfer2}) and eq.~(\ref{eq:interfer1}) are subtracted due to the overall $\sigma$ in $f_{{\cal B} 7}^{}$, and the interference terms contribute in the same direction. Therefore, as long as $f_{\gamma}^{\lambda}$ (i.e the $HZ\gamma$ coupling) is small, the effects of it appear larger in $F_{{\cal B} 7}^{}$ than in $F_{{\cal C} 8}^{}$. 
Notice that this discussion is independent on explicit choices for $\lambda$ and $\lambda^{\prime}_{}$. The above finding is true not only in $F_{{\cal B} 7}^{}$ but also in the other coefficients which have the overall $\sigma$ in $f_{{i}a}^{}$. These coefficients are indicated by the symbol $\circ$ in the column "$HZ\gamma$" in Table~\ref{table:symproperty}. \\

In Figure~\ref{figure:ZHplot2}, the coefficients $F_{{\cal B} 8}^{}$ (left panel) and $F_{{\cal C} 7}^{}$ (right panel) divided by $F_{{\cal A} 1}^{}$ are shown in the same manner as Figure~\ref{figure:ZHplot1}. 
$\mathrm{CP\widetilde{T}}$ invariance requires these 2 coefficients to be identically zero, hence observation of a non-zero value indicates the existence of re-scattering effects. 
Re-scattering effects can be approximately included by allowing imaginary parts in the form factors~\cite{Hagiwara:1986vm}. 
In each panel, $\Box$ points give predictions for $h_3^Z=0+0.1i$, and $\times$ points give those for $h_3^{\gamma}=0+0.1i$. \\

In Figure~~\ref{figure:ZHplot3}, the coefficients $F_{{\cal A} 3}^{}$ (left panel), $F_{{\cal D} 1}^{}$ (middle-left panel), $-F_{{\cal B} 5}^{}$ (middle-right panel) and $-F_{{\cal C} 4}^{}$ (right panel) divided by $F_{{\cal A} 1}^{}$ are shown in the same manner as Figure~\ref{figure:ZHplot1}. These 4 coefficients are doubly constrained to be identically zero by CP invariance {\it and} $\mathrm{CP\widetilde{T}}$ invariance: they are strictly zero if CP is conserved, even if $\mathrm{CP\widetilde{T}}$ is violated, for instance. Observation of a non-zero value indicates CP violation {\it and} the existence of re-scattering effects. In each panel, $\Box$ points give predictions for the CP-odd from factor $\widetilde{h}_4^Z=0+0.1i$, and $\times$ points give those for the CP-odd form factor $\widetilde{h}_4^{\gamma}=0+0.1i$.

\section{Polarisation of the $W$ boson}\label{sec:polWboson}

In this section, we analyse the density matrices of the $W^{+}_{}$ and $W^{-}_{}$ in the process $pp \to W^{\pm}_{}H$ following the same procedure as Section~\ref{sec:densitymatrix}. 
Because (1) a determination of the density matrix of the $W$ boson is difficult in $pp$ collisions when the $W$ boson decays into a charged lepton and a neutrino and (2) we cannot distinguish $W^+_{}$ from $W^-_{}$ in view of the difficulty of flavor identification of both $W^+_{}$ and $W^-_{}$ decay products when the $W$ boson decays into two quarks, we consider the process $pp \to W^{\pm}_{}H$ ($W^{\pm}_{} \to jj$) as the sum of the process $pp \to W^+_{}H$ ($W^+_{} \to jj$) and the process $pp \to W^-_{}H$ ($W^-_{} \to jj$). 
Just as the process $pp\to ZH$, only the $15$ coefficients among the $36$ coefficients of the 4 different differential angular distributions of the dijets can be non-zero. However only 9 coefficients among these $15$ coefficients are actually measurable.

\subsection{Polarisation density matrices of the $W^+_{}$ and $W^-_{}$}\label{sec:densitymatrixW}

First of all, we define the density matrices of the $W^+_{}$ and $W^-_{}$. We consider the sub-processes
\begin{subequations}
\begin{align}
u(\sigma) + \bar{d}( -\sigma ) & \to W^+_{}(\lambda) + H; \nonumber \\
W^+_{}(\lambda) & \to u(\tau) + \bar{d}(-\tau),
\end{align}
and
\begin{align}
d(\sigma) + \bar{u}( -\sigma ) & \to W^-_{}(\lambda) + H; \nonumber \\
W^-_{}(\lambda) & \to d(\tau) + \bar{u}(-\tau),
\end{align}
\end{subequations}
where the helicity of each particle is shown in parenthesis. We neglect the masses of the quark and the antiquark from the $W$ decay, hence the helicity of the antiquark is always opposite to that of the quark. 
We prepare the 4 full helicity amplitudes ${\cal T}(u\bar{d})$, ${\cal T}(d\bar{u})$, ${\cal T}(\bar{d}u)$ and ${\cal T}(\bar{u}d)$ in which the production amplitude is evaluated in the $u\bar{d}$ c.m. frame, the $d\bar{u}$ c.m. frame, the $\bar{d}u$ c.m. frame and the $\bar{u}d$ c.m. frame, respectively. 
Recall that these c.m. frames are shown in Figure~\ref{figure:frames} and the production amplitudes are given in eqs.~(\ref{eq:WpampAll}) and (\ref{eq:WmampAll}). 
These 4 full helicity amplitudes are given by
\begin{subequations}
\begin{align}
{\cal T}(u\bar{d}) & =  P_W^{} \sum_{\lambda=\pm,0}^{} {\cal M}_{-}^{\lambda}(u \bar{d})\ D_{\lambda}^{},\label{eq:fullampW1} \\
{\cal T}(\bar{d}u) & =  P_W^{} \sum_{\lambda=\pm,0}^{} {\cal M}_{-}^{\lambda}(\bar{d} u)\ D_{\lambda}^{},\label{eq:fullampW2} \\
{\cal T}(d\bar{u}) & =  P_W^{} \sum_{\lambda=\pm,0}^{} {\cal M}_{-}^{\lambda}(d \bar{u})\  \widetilde{D}_{\lambda}^{},\label{eq:fullampW3} \\
{\cal T}(\bar{u}d) & =  P_W^{} \sum_{\lambda=\pm,0}^{} {\cal M}_{-}^{\lambda}(\bar{u}d)\ \widetilde{D}_{\lambda}^{},\label{eq:fullampW4} 
\end{align}
\end{subequations}
where 
\begin{align}
P_W^{} =  ( Q^2_{} - m_W^2 + i m_W^{} \Gamma_W^{} )^{-1}_{}
\end{align}
denotes the propagator factor of the $W$ boson, $D^{}_{\lambda}$ is the helicity amplitude for the decay process $W^+_{} \to u \bar{d}$ and $\widetilde{D}^{}_{\lambda}$ is the helicity amplitude for the decay process $W^-_{} \to d \bar{u}$:
\begin{subequations}
\begin{align}
D^{}_{\lambda} & = \frac{g}{\sqrt{2}} V_{ud}^{} m_W^{}\ d^-_{\lambda}, \\
\widetilde{D}^{}_{\lambda} & = \frac{g}{\sqrt{2}} (V_{ud}^{})^*_{} m_W^{}\ d^-_{\lambda},
\end{align}
\end{subequations}
where $d^-_{\lambda}$ is given in eq.~(\ref{eq:decayamp}). 
The decay amplitudes are evaluated in the following four-momentum frame:
\begin{align}
W^+_{}\ (W^-_{}):&\ \ \bigl( m_W^{},0, 0, 0 \bigr) \nonumber \\
u \ \ (d):&\ \  \frac{m_W^{}}{2}\bigl( 1,\ \sin{\widehat{\theta}} \cos{\widehat{\phi}},\ \sin{\widehat{\theta}} \sin{\widehat{\phi}},\ \cos{\widehat{\theta}} \bigr) \nonumber \\
\bar{d}\ \ (\bar{u}) :&\ \  \frac{m_W^{}}{2}\bigl( 1,\ -\sin{\widehat{\theta}} \cos{\widehat{\phi}},\ -\sin{\widehat{\theta}} \sin{\widehat{\phi}},\ -\cos{\widehat{\theta}} \bigr). \label{eq:decaykinematics-W}
\end{align}
Then we obtain
\begin{subequations}
\begin{align}
\bigl| {\cal T}(u\bar{d}) \bigr|^2_{}
 & = \bigl|P_W^{} m_W^{} g V_{ud}^{} / \sqrt{2} \bigr|^2_{}
\sum_{\lambda^{\prime}_{}, \lambda} \bigl( d_{\lambda^{\prime}}^{-} \bigr)^*_{} \rho_{}^{\lambda^{\prime} \lambda}(u\bar{d})\ d_{\lambda}^{-}, \label{eq:ampsquaredW1} \\
\bigl| {\cal T}(\bar{d}u) \bigr|^2_{}
 & = \bigl|P_W^{} m_W^{} g V_{ud}^{} / \sqrt{2} \bigr|^2_{}
\sum_{\lambda^{\prime}_{}, \lambda} \bigl( d_{\lambda^{\prime}}^{-} \bigr)^*_{} \rho_{}^{\lambda^{\prime} \lambda}(\bar{d}u)\ d_{\lambda}^{-}, \label{eq:ampsquaredW2} \\
\bigl| {\cal T}(d\bar{u}) \bigr|^2_{}
 & = \bigl|P_W^{} m_W^{} g V_{ud}^{} / \sqrt{2} \bigr|^2_{}
\sum_{\lambda^{\prime}_{}, \lambda} \bigl( d_{\lambda^{\prime}}^{-} \bigr)^*_{} \rho_{}^{\lambda^{\prime} \lambda}(d\bar{u})\ d_{\lambda}^{-}, \label{eq:ampsquaredW3} \\
\bigl| {\cal T}(\bar{u}d) \bigr|^2_{}
 & = \bigl|P_W^{} m_W^{} g V_{ud}^{} / \sqrt{2} \bigr|^2_{}
\sum_{\lambda^{\prime}_{}, \lambda} \bigl( d_{\lambda^{\prime}}^{-} \bigr)^*_{} \rho_{}^{\lambda^{\prime} \lambda}(\bar{u}d)\ d_{\lambda}^{-}, \label{eq:ampsquaredW4}
\end{align}
\end{subequations}
where 
\begin{subequations}\label{eq:densityMW12}
\begin{align}
\rho_{}^{\lambda^{\prime} \lambda}(u\bar{d}) & = 
 \bigl\{ {\cal M}_{-}^{\lambda^{\prime}_{}}(u \bar{d}) \bigr\}^{\ast}_{}
 {\cal M}_{-}^{\lambda}(u \bar{d}),\label{eq:densityMW1} \\
\rho_{}^{\lambda^{\prime} \lambda}(\bar{d}u) & = 
 \bigl\{ {\cal M}_{-}^{\lambda^{\prime}_{}}(\bar{d}u) \bigr\}^{\ast}_{}
 {\cal M}_{-}^{\lambda}(\bar{d}u),\label{eq:densityMW2} 
\end{align}
\end{subequations}
represent the elements of the density matrix in the helicity basis of the $W^+_{}$ in the $u\bar{d}$ c.m. frame and those in the $\bar{d}u$ c.m. frame, respectively, and
\begin{subequations}\label{eq:densityMW34}
\begin{align}
\rho_{}^{\lambda^{\prime} \lambda}(d \bar{u}) & = 
 \bigl\{ {\cal M}_{-}^{\lambda^{\prime}_{}}(d \bar{u}) \bigr\}^{\ast}_{}
 {\cal M}_{-}^{\lambda}(d \bar{u}),\label{eq:densityMW3} \\
\rho_{}^{\lambda^{\prime} \lambda}(\bar{u} d) & = 
 \bigl\{ {\cal M}_{-}^{\lambda^{\prime}_{}}(\bar{u} d) \bigr\}^{\ast}_{}
 {\cal M}_{-}^{\lambda}(\bar{u} d),\label{eq:densityMW4} 
\end{align}
\end{subequations}
represent the elements of the density matrix in the helicity basis of the $W^-_{}$ in the $d\bar{u}$ c.m. frame and those in the $\bar{u}d$ c.m. frame, respectively. 
By looking at the amplitudes in eqs.~(\ref{eq:WpampAll}) and (\ref{eq:WmampAll}), it is easy to find the relations
\begin{subequations}\label{eq:relationdensitymatrix}
\begin{align}
\rho_{}^{\lambda^{\prime} \lambda}(u\bar{d}) & = \rho_{}^{\lambda^{\prime} \lambda}(d\bar{u}), \\
\rho_{}^{\lambda^{\prime} \lambda}(\bar{d}u) & = \rho_{}^{\lambda^{\prime} \lambda}(\bar{u}d). 
\end{align}
\end{subequations}
The relations indicate that the $W^+_{}$ and $W^-_{}$ are always in the same state of polarisation. This fact makes our analysis in the next section simpler, because we do not have to distinguish the $W^+_{}$ from the $W^-_{}$ with regard to states of polarisation.

\subsection{Decay angular distributions of the polarised $W^+_{}$ and $W^-_{}$}\label{sec:corsspolW}

The complete differential cross section for the process $pp \to W^+_{}H$ followed by $W^+_{} \to jj$ in the narrow width approximation can be expressed in terms of the density matrices in eq.~(\ref{eq:densityMW12}) as
\begin{align}
\frac{d\sigma^{W^+_{}}_{}}{d\hat{s}\ dy\ d\cos{\theta}\ d\cos{\widehat{\theta}}\ d\widehat{\phi} }
= & \frac{m_W^{}k}{ 8192 \pi^3_{} \Gamma_W^{} s \hat{s}^{\frac{3}{2}}_{}} 
g^2_{} \Bigl(\sum_{u, d} \bigl| V_{ud}^{} \bigr|^2_{} \Bigr)\nonumber \\
& \times \sum_{u,d} \biggl[ u(x_1^{}) \bar{d}(x_2^{})\ d_{}^{- \dagger} \rho_{}^{}(u\bar{d}) d_{}^{-}
+
\bar{d}(x_1^{}) u(x_2^{})\ d_{}^{- \dagger} \rho_{}^{}(\bar{d}u) d_{}^{-} \biggr],\label{eq:differentialWH1}
\end{align}
and that for the process $pp \to W^-_{}H$ followed by $W^-_{} \to jj$ can be expressed in terms of the density matrices in eq.~(\ref{eq:densityMW34}) as
\begin{align}
\frac{d\sigma^{W^-_{}}_{}}{d\hat{s}\ dy\ d\cos{\theta}\ d\cos{\widehat{\theta}}\ d\widehat{\phi} }
= & \frac{m_W^{}k}{ 8192 \pi^3_{} \Gamma_W^{} s \hat{s}^{\frac{3}{2}}_{}} 
g^2_{} \Bigl(\sum_{u, d} \bigl| V_{ud}^{} \bigr|^2_{} \Bigr)\nonumber \\
& \times \sum_{u,d} \biggl[ d(x_1^{}) \bar{u}(x_2^{})\ d_{}^{- \dagger} \rho_{}^{}(d\bar{u}) d_{}^{-}
+
\bar{u}(x_1^{}) d(x_2^{})\ d_{}^{- \dagger} \rho_{}^{}(\bar{u}d) d_{}^{-} \biggr], \label{eq:differentialWH2}
\end{align}
where the $3\times 3$ matrix form in eq.~(\ref{eq:matrixform}) is used. For the definition of variables, see below eq.~(\ref{eq:differentialZH}). 
As the quark flavors $u$ and $d$ summed in the above equations, we consider all the quark flavors but the top quark. As a result, the unitarity of the CKM matrix gives $\sum_{u, d} | V_{ud}^{} |^2_{} = 2$. By using the relations in eq.~(\ref{eq:relationdensitymatrix}), we obtain the complete differential cross section for the process $pp \to W^{\pm}_{}H$ followed by $W^{\pm}_{} \to jj$ as the sum of the above two cross sections in the following compact form:
\begin{align}
\frac{d\sigma^{W^+_{}+W^-_{}}_{}}{d\hat{s}\ dy\ d\cos{\theta}\ d\cos{\widehat{\theta}}\ d\widehat{\phi} }
 & =  \frac{m_W^{}k}{ 4096 \pi^3_{} \Gamma_W^{} s \hat{s}^{\frac{3}{2}}_{}} 
g^2_{} \nonumber \\
\times \sum_{u,d} \biggl[  \Bigl\{ u(x_1^{}) & \bar{d}(x_2^{}) + d(x_1^{}) \bar{u}(x_2^{}) \Bigr\}\ d_{}^{- \dagger} \rho(u\bar{d}) d_{}^{-}
+
\Bigl\{ \bar{u}(x_1^{}) d(x_2^{}) + \bar{d}(x_1^{}) u(x_2^{}) \Bigr\}\ d_{}^{- \dagger} \rho(\bar{d}u) d_{}^{-}  \biggr].\label{eq:differentialWH}
\end{align}
As in Section~\ref{sec:crossZboson}, we derive the differential cross section with respect to $\hat{s}$, $\cos{\widehat{\theta}}$ and $\widehat{\phi}$ by integrating over $y$ and $\cos{\theta}$ in eq.~(\ref{eq:differentialWH}). 
It is straightforward to confirm that the integration over $y$ and $\cos{\theta}$ in the 4 different approaches summarised in Table~\ref{table:integration} can be performed in the same manner as eqs.~(\ref{eq:PartdifferentialZH1}), (\ref{eq:PartdifferentialZH2}), (\ref{eq:PartdifferentialZH3}) and (\ref{eq:PartdifferentialZH4}). 
The resulting differential cross sections are
\begin{subequations}
\begin{align}
\frac{d\sigma^{W^+_{}+W^-_{}}}{d\hat{s}\ d\cos{\widehat{\theta}}\ d\widehat{\phi} }\biggr|_{{\cal A}}^{}
  = &
{\cal T}^W_{} \int_{0}^{y_{\mathrm{cut}}^{}} dy\ 2 \Bigl[  u(x_1^{}) \bar{d}(x_2^{}) + d(x_1^{}) \bar{u}(x_2^{}) +  \bar{u}(x_1^{}) d(x_2^{}) + \bar{d}(x_1^{}) u(x_2^{}) \Bigr] \nonumber \\
& \times
d_{}^{- \dagger} 
\left(
\begin{array}{ccc}
c_1^{} | \hat{N}_{}^+ |^2_{} & \frac{1}{2} c_2^{} ( \hat{N}_{}^{+} )^*_{} \hat{N}_{}^{-}  & 0 \\
\frac{1}{2} c_2^{} \hat{N}_{}^{+} ( \hat{N}_{}^{-} )^*_{}  & c_1^{} | \hat{N}_{}^- |^2_{} & 0  \\
0 & 0 & c_2^{} | \hat{N}_{}^0 |^2_{} \\
\end{array}
\right)
d_{}^{-}
,\label{eq:PartdifferentialWH1}\\
\frac{d\sigma^{W^+_{}+W^-_{}}}{d\hat{s}\ d\cos{\widehat{\theta}}\ d\widehat{\phi} }\biggr|_{{\cal B}}^{}
  = &
{\cal T}^W_{} \int_{0}^{y_{\mathrm{cut}}^{}} dy\ 2 \Bigl[  u(x_1^{}) \bar{d}(x_2^{}) + d(x_1^{}) \bar{u}(x_2^{}) -  \bar{u}(x_1^{}) d(x_2^{}) - \bar{d}(x_1^{}) u(x_2^{}) \Bigr] \nonumber \\
& \times
d_{}^{- \dagger} 
\left(
\begin{array}{ccc}
0 & 0  & - c_3^{} ( \hat{N}_{}^{+} )^*_{} \hat{N}_{}^{0} \\
0  & 0 & - c_3^{} ( \hat{N}_{}^{-} )^*_{} \hat{N}_{}^{0}  \\
- c_3^{} \hat{N}_{}^{+} ( \hat{N}_{}^{0} )^*_{} & - c_3^{} \hat{N}_{}^{-} ( \hat{N}_{}^{0} )^*_{} & 0 \\
\end{array}
\right)
d_{}^{-},\label{eq:PartdifferentialWH2} \\
\frac{d\sigma^{W^+_{}+W^-_{}}}{d\hat{s}\ d\cos{\widehat{\theta}}\ d\widehat{\phi} }\biggr|_{{\cal C}}^{}
  = &
{\cal T}^W_{} \int_{0}^{y_{\mathrm{cut}}^{}} dy\ 2 \Bigl[  u(x_1^{}) \bar{d}(x_2^{}) + d(x_1^{}) \bar{u}(x_2^{}) +  \bar{u}(x_1^{}) d(x_2^{}) + \bar{d}(x_1^{}) u(x_2^{}) \Bigr] \nonumber \\
& \times
d_{}^{- \dagger} 
\left(
\begin{array}{ccc}
0 & 0  & c_4^{} ( \hat{N}_{}^{+} )^*_{} \hat{N}_{}^{0}  \\
0 & 0 & - c_4^{} ( \hat{N}_{}^{-} )^*_{} \hat{N}_{}^{0}   \\
c_4^{} \hat{N}_{}^{+} ( \hat{N}_{}^{0} )^*_{}  & - c_4^{} \hat{N}_{}^{-} ( \hat{N}_{}^{0} )^*_{}  & 0 \\
\end{array}
\right)
d_{}^{-},
\label{eq:PartdifferentialWH3}\\
\frac{d\sigma^{W^+_{}+W^-_{}}}{d\hat{s}\ d\cos{\widehat{\theta}}\ d\widehat{\phi} }\biggr|_{{\cal D}}^{}
  = &
{\cal T}^W_{} \int_{0}^{y_{\mathrm{cut}}^{}} dy\ 2 \Bigl[  u(x_1^{}) \bar{d}(x_2^{}) + d(x_1^{}) \bar{u}(x_2^{}) -  \bar{u}(x_1^{}) d(x_2^{}) - \bar{d}(x_1^{}) u(x_2^{}) \Bigr] \nonumber \\
& \times
d_{}^{- \dagger} 
\left(
\begin{array}{ccc}
- c_5^{}  | \hat{N}_{}^+ |^2_{} & 0  & 0 \\
0  &  c_5^{} | \hat{N}_{}^- |^2_{} & 0  \\
0 & 0 & 0 \\
\end{array}
\right)
d_{}^{-},\label{eq:PartdifferentialWH4}
\end{align}
\end{subequations}
where
\begin{align}
{\cal T}^W_{} =\frac{m_W^{}k}{ 4096 \pi^3_{} \Gamma_W^{} s \hat{s}^{\frac{3}{2}}_{}} g^2_{} \sum_{u, d} \bigl| V_{ud}^{} \bigr|^2_{},
\end{align}
the constant values $c_i^{}$ $(i=1,2,3,4,5)$ are summarised in Table~\ref{table:coefficients}, and $\hat{N}^{\lambda=\pm,0}_{}$ are defined in eq.~(\ref{eq:Wsubamplitudes}). By comparing the above 4 differential cross sections with eq.~(\ref{eq:diffcrossgeneral}), we obtain the 36 coefficients $F_{ia}^{W}$ $(i={\cal A, B, C, D})$ $(a=1,2,\cdots,9)$ in total:
\begin{subequations}\label{eq:observablesppWH}
\begin{align}
F_{ {\cal A} ({\cal C}) a}^{W}
& =
{\cal T}^W_{} \int_{0}^{y_{\mathrm{cut}}^{}} dy\ 2 \Bigl[  u(x_1^{}) \bar{d}(x_2^{}) + d(x_1^{}) \bar{u}(x_2^{}) +  \bar{u}(x_1^{}) d(x_2^{}) + \bar{d}(x_1^{}) u(x_2^{}) \Bigr] f_{ {\cal A} ({\cal C}) a}^{W}, \\
F_{ {\cal B} ({\cal D}) a}^{W}
& =
{\cal T}^W_{} \int_{0}^{y_{\mathrm{cut}}^{}} dy\ 2 \Bigl[  u(x_1^{}) \bar{d}(x_2^{}) + d(x_1^{}) \bar{u}(x_2^{}) -  \bar{u}(x_1^{}) d(x_2^{}) - \bar{d}(x_1^{}) u(x_2^{}) \Bigr] f_{ {\cal B} ({\cal D}) a}^{W},
\end{align}
\end{subequations}
where $f_{ia}^{W}$ $(i={\cal A, B, C, D})$ $(a=1,2,\cdots,9)$ are obtained by the following replacements in those coefficients of the process $pp \to ZH$ in eq.~(\ref{eq:observables-small}):
\begin{align}
\hat{M}_{\sigma}^{\lambda} \to \hat{N}_{}^{\lambda}, \ \ 
\sigma \to -1,\ \ 
\tau \to -1.
\end{align}
Table~\ref{table:symproperty} corresponding to the process $pp \to W^{\pm}_{}H$ is obtained by the following simple replacements in Table~\ref{table:symproperty}:
\begin{align}
\hat{M}_{\sigma}^{\lambda} \to \hat{N}_{}^{\lambda}, \ \ 
F_{ia}^{} \to F_{ia}^{W}. 
\end{align}
Just as the process $pp\to ZH$, there are in total $15$ coefficients which can be non-zero. 
Observation of the 6 coefficients $F_{{\cal D} 3}^{W}$, $F_{{\cal A} 3}^{W}$, $F_{{\cal B} 4}^{W}$, $F_{{\cal C} 4}^{W}$, $F_{{\cal B} 7}^{W}$ and $F_{{\cal C} 7}^{W}$ has difficulty, since it requires the charge (or flavor) identification of the parent quark of the jet. Only the remaining 9 coefficients $F_{{\cal A} 1}^{W}$, $F_{{\cal A} 2}^{W}$, $F_{{\cal D} 1}^{W}$, $F_{{\cal C} 5}^{W}$, $F_{{\cal B} 5}^{W}$, $F_{{\cal A} 6}^{W}$, $F_{{\cal C} 8}^{W}$, $F_{{\cal B} 8}^{W}$ and $F_{{\cal A} 9}^{W}$ are, therefore, actually measurable. We emphasise that these 9 coefficients are necessary and sufficient to determine all of the 9 independent combinations of the density matrix elements.

\subsection{Influences of non-standard $HWW$ interaction}\label{sec:angcoeff-W}

\begin{figure}[t]
\centering
\includegraphics[scale=0.45]{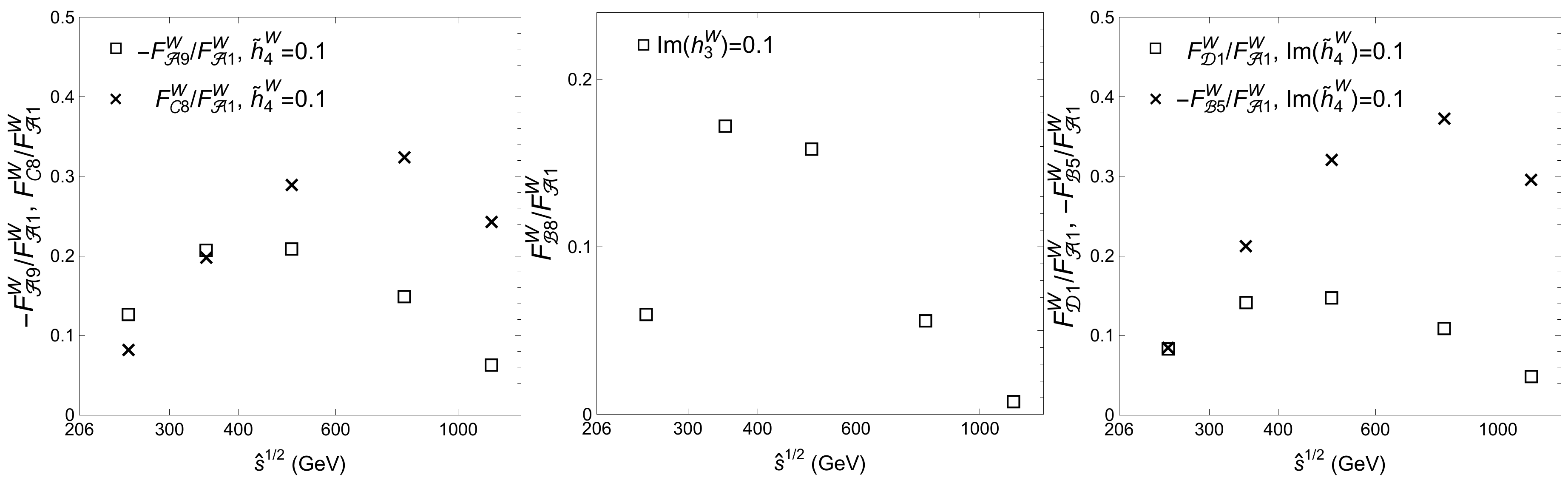}
\caption{\small 
In the left panel, the 2 coefficients $-F_{{\cal A} 9}^{W}$ ($\Box$) and $F_{{\cal C} 8}^{W}$ ($\times$) divided by $F_{{\cal A} 1}^{W}$ are shown for $\widetilde{h}_4^W=0.1$. These 2 coefficients are constrained to be identically zero by CP invariance. 
In the middle panel, the coefficient $F_{{\cal B} 8}^{W}$ divided by $F_{{\cal A} 1}^{W}$ are shown for $h_3^W=0+0.1i$. This coefficient is constrained to be identically zero by $\mathrm{CP\widetilde{T}}$ invariance. 
In the right panel, the 2 coefficients $F_{{\cal D} 1}^{W}$ ($\Box$) and $-F_{{\cal B} 5}^{W}$ ($\times$) divided by $F_{{\cal A} 1}^{W}$ are shown for $\widetilde{h}_4^W=0+0.1i$.  
These 2 coefficients are doubly constrained to be identically zero by CP invariance {\it and} $\mathrm{CP\widetilde{T}}$ invariance. 
For each result, the values in the 5 bins, from left to right, are obtained after integration over $\hat{s}$ in the regions $(m_W^{}+m_H^{}) < \hat{s}^{1/2}_{} < 300$, $300 < \hat{s}^{1/2}_{} < 400$, $400 < \hat{s}^{1/2}_{} < 600$, $600 < \hat{s}^{1/2}_{} < 1000$ and $1000 < \hat{s}^{1/2}_{} < 14000$ in units of GeV. }
\label{figure:WHplot}
\end{figure}

In this Section, with the same motivation as Section~\ref{sec:angcoeff}, we focus on the measurable coefficients which are strictly zero or small in the SM, and study the influences of the non-standard $HWW$ coupling. 
In the left panel of Figure~\ref{figure:WHplot}, the 2 coefficients $-F_{{\cal A} 9}^{W}$ ($\Box$) and $F_{{\cal C} 8}^{W}$ ($\times$) divided by $F_{{\cal A} 1}^{W}$ are shown for the CP-odd form factor $\widetilde{h}_4^W=0.1$. CP invariance requires these 2 coefficients to be identically zero; observation of a non-zero value signals CP violation. 
In the middle panel of Figure~\ref{figure:WHplot}, the coefficient $F_{{\cal B} 8}^{W}$ divided by $F_{{\cal A} 1}^{W}$ is shown for the CP-even form factor $h_3^W=0+0.1i$. $\mathrm{CP\widetilde{T}}$ invariance requires this coefficient to be identically zero; observation of a non-zero value indicates the existence of re-scattering effects. 
In the right panel of Figure~\ref{figure:WHplot}, the 2 coefficients $F_{{\cal D} 1}^{W}$ ($\Box$) and $-F_{{\cal B} 5}^{W}$ ($\times$) divided by $F_{{\cal A} 1}^{W}$ are shown for the CP-odd form factor $\widetilde{h}_4^W=0+0.1i$. 
These 2 coefficients are doubly constrained to be identically zero by CP invariance {\it and} $\mathrm{CP\widetilde{T}}$ invariance; observation of a non-zero value indicates CP violation {\it and} the existence of re-scattering effects. 
For each result, the values in the 5 bins, from left to right, are obtained after integration over $\hat{s}$ in the regions $(m_W^{}+m_H^{}) < \hat{s}^{1/2}_{} < 300$, $300 < \hat{s}^{1/2}_{} < 400$, $400 < \hat{s}^{1/2}_{} < 600$, $600 < \hat{s}^{1/2}_{} < 1000$ and $1000 < \hat{s}^{1/2}_{} < 14000$ in units of GeV. \\

We discuss the differences between the coefficients $F_{ia}^{}$ in the process $pp\to ZH$ and those $F_{ia}^{W}$ in the process $pp\to W^{\pm}_{}H$. 
From the comparisons between the results for the non-standard $HZZ$ coupling in Figures~\ref{figure:ZHplot1}, \ref{figure:ZHplot2} and \ref{figure:ZHplot3} and the results for the non-standard $HWW$ coupling in Figure~\ref{figure:WHplot}, we notice that $F_{{\cal A} 9}^{W}$ and $F_{{\cal C} 8}^{W}$ are comparable with $F_{{\cal A} 9}^{}$ and $F_{{\cal C} 8}^{}$, respectively, 
while $F_{{\cal B} 8}^{W}$, $F_{{\cal D} 1}^{W}$ and $F_{{\cal B} 5}^{W}$ are consistently larger than $F_{{\cal B} 8}^{}$, $F_{{\cal D} 1}^{}$ and $F_{{\cal B} 5}^{}$, respectively. 
As we have discussed at the last paragraph of Section~\ref{sec:helamp}, the $Z$ boson is in a partially polarised state, while the $W^+_{}$ and $W^-_{}$ are in a completely polarised state. The degree of polarisation affects the magnitudes of the coefficients, hence it is expected that the coefficients $F_{ia}^{W}$ are always equal or larger than the coefficients $F_{ia}^{}$. The existence of the overall $\sigma$ in $F_{{\cal B} 8}^{}$, $F_{{\cal D} 1}^{}$ and $F_{{\cal B} 5}^{}$ (see eq.~(\ref{eq:observables-small})) indicates that these coefficients are weakened according to the degree of polarisation of the $Z$ boson. These coefficients are denoted by the symbol $\circ$ in the last column in Table~\ref{table:symproperty}.

\section{Summary}\label{sec:summary}

The measurements of the Higgs boson couplings to the SM particles are essential tests of the SM. 
The $Z$ boson in the process $pp\to ZH$ and the $W^+_{}$ and $W^-_{}$ in the process $pp \to W^{\pm}_{}H$ can be in polarised states, and it would be possible to study the Higgs boson couplings to the weak bosons ($HZZ$, $HZ\gamma$ and $HWW$) in detail from a careful analysis of these states of polarisation. 
A density matrix contains the complete information about a state of polarisation, and all of the elements of the density matrix should be made use of in such a careful analysis. 
In this paper, such an analysis approach has been presented. \\

Determination of the density matrix of the $Z$ ($W$) boson requires measurements of the angular distributions of the $Z$ ($W$) decay products. In $pp$ collisions, this is difficult when the $W$ boson decays into a charged lepton and a neutrino. When the $W$ boson decays into two quarks, we cannot distinguish $W^+_{}$ from $W^-_{}$ in view of the difficulty of flavor identification of both $W^+_{}$ and $W^-_{}$ decay products. 
Therefore, we have considered the process $pp \to W^{\pm}_{}H$ ($W^{\pm}_{} \to jj$) as the sum of the process $pp \to W^+_{}H$ ($W^+_{} \to jj$) and the process $pp \to W^-_{}H$ ($W^-_{} \to jj$). 
We have found that the $W^+_{}$ and $W^-_{}$ are always in the same state of polarisation (i.e. the density matrix of the $W^+_{}$ is always the same as that of $W^-_{}$). 
By using this fact, we have developed an analysis approach which can be applied both to $pp \to ZH$ ($Z \to f\bar{f}$) and to $pp \to W^{\pm}_{}H$ ($W^{\pm}_{} \to jj$) in the same manner. \\

We have derived the 4 different differential cross sections with respect to $\hat{s}$ (the invariant mass squared of the $Z$ ($W$) boson and the Higgs boson) and $\cos{\widehat{\theta}}$ and $\widehat{\phi}$ (the $Z$ ($W$) decay angles), by integrating the complete differential cross section over the other phase space variables in the 4 different approaches. 
Among the 36 coefficients ($=9 \times 4$) of these 4 differential angular distributions, only 15 coefficients can be non-zero (these coefficients are summarised in Table~\ref{table:symproperty}). 
These 15 coefficients are written in terms of the elements of the density matrices, and there exist 9 independent combinations of the elements of the density matrices. 
Observation of the 9 coefficients among the 15 coefficients does not require the charge (or flavor) identification of the $Z$ ($W$) decay products. 
In the analysis of the $W^+_{}$ and $W^-_{}$, only these 9 coefficients are measurable. We have found that these 9 coefficients are necessary and sufficient to determine all of the 9 independent combinations of the density matrix elements (i.e. one coefficient corresponds to one combination). 
We have clarified the restrictions on the 15 coefficients imposed by the CP and $\mathrm{CP\widetilde{T}}$ symmetries. 
Some of the coefficients are required to be identically zero by CP invariance; observation of a non-zero value in these coefficients signals CP violation. Similarly, some of the coefficients are required to be identically zero by $\mathrm{CP\widetilde{T}}$ invariance; observation of a non-zero value in these coefficients indicates the existence of re-scattering effects. 
These coefficients are in particular interesting as observables at the LHC, since observation of a non-zero or large value in these coefficients immediately signals the existence of non-standard $HZZ$, $HZ\gamma$ and/or $HWW$ interactions.

\section*{Acknowledgments}

I would like to thank Kaoru Hagiwara for many discussions. I am also grateful to members of CP3, Universit\'e Catholique de Louvain for their hospitality. 
My work is supported by the Alexander von Humboldt Foundation.

\small

\bibliographystyle{JHEP}
\nocite{*}
\bibliography{zh_bib}

\end{document}